%% file: main.tex
\documentclass[
 preprint, 
superscriptaddress,
 amsmath,amssymb,
 aps, physrev,
]{revtex4-2}

\input{commands}

\usepackage{amsmath}
\usepackage{amsfonts}
\usepackage{amsthm}
\usepackage{subcaption}
\usepackage{natbib}        
\usepackage{algorithm}
\usepackage{algpseudocode}
\usepackage[hidelinks]{hyperref}
\usepackage{graphicx}

\begin{document}

\preprint{APS/123-QED}

\title{Consistent Projection of Langevin Dynamics: Preserving Thermodynamics and Kinetics in Coarse-Grained Models
}%

\author{Vahid Nateghi}
 \affiliation{
 Max-Planck-Institute for Dynamics of Complex Technical Systems, Magdeburg, Germany}%

\author{Lara Neureither}%
 \affiliation{Institute of Mathematics,
 Brandenburgische Technische Universität Cottbus-Senftenberg, Cottbus, Germany}%

\author{Selma Moqvist}%
 \affiliation{Department of Computer Science and Engineering,
 Chalmers University of Technology and University of Gothenburg, Gothenburg, Sweden}%

 \author{Carsten Hartmann}%
 \affiliation{
 Brandenburgische Technische Universität Cottbus-Senftenberg, Cottbus, Germany}%

\author{Simon Olsson}%
 \affiliation{Department of Computer Science and Engineering,
 Chalmers University of Technology and University of Gothenburg, Gothenburg, Sweden}%

\author{Feliks Nüske}\email{nueske@mpi-magdeburg.mpg.de}
 \affiliation{
 Max-Planck-Institute for Dynamics of Complex Technical Systems, Magdeburg, Germany}%

\date{\today}

\begin{abstract}
Coarse graining (CG) is an important task for efficient modeling and simulation of complex multi-scale systems, such as the conformational dynamics of biomolecules. This work presents a projection-based coarse-graining formalism for general underdamped Langevin dynamics. Following the Zwanzig projection approach, we derive a closed-form expression for the coarse grained dynamics. In addition, we show how the generator Extended Dynamic Mode Decomposition (gEDMD) method, which was developed in the context of Koopman operator methods, can be used to model the CG dynamics and evaluate its kinetic properties, such as transition timescales. Finally, we combine our approach with thermodynamic interpolation (TI), a generative approach to transform samples between thermodynamic conditions, to extend the scope of the approach across thermodynamic states without repeated numerical simulations. Using a two-dimensional model system, we demonstrate that the proposed method allows to accurately capture the thermodynamic and kinetic properties of the full-space model.
\end{abstract}

\maketitle


\section{Introduction}

The task of finding a simpler representation for a complex dynamical system, while preserving essential features of the full system, is referred to as \textbf{coarse graining (CG)} or model reduction~\cite{das_low-dimensional_2006,clementi_coarse-grained_2008,benner_survey_2015}. CG models are central across many scientific disciplines because they enable interpretability and simulation of large-scale systems at greatly reduced computational cost. In molecular dynamics, CG modeling has long been used to study biomolecular systems whose size and timescales are inaccessible to all-atom simulations~\cite{jin_bottom-up_2022}. Since simulations of biomolecules often exhibit rare events and meta stable states, constructing high-quality CG models that can reduce the computational burden is of particular importance.

The quality of a CG model depends both on the function that is used to map a full-state configuration to its reduced representation (CG map), as well as the dynamical model defined on the reduced space (CG dynamics). While numerous methods exist for defining effective CG maps~\cite{rohrdanz_discovering_2013,sidky_machine_2020,gkeka_machine_2020}, the accurate construction of CG dynamics remains more challenging. Different metrics have been suggested to assess the quality of a CG model. For equilibrium systems, reproducing the marginal steady-state distribution is a common criterion of quality as formalized by the multi-scale CG method~\cite{noid_multiscale_2008} and its machine-learning-based variants~\cite{wang_machine_2019,charron_navigating_2025}. However, ensuring that the CG dynamics also reproduces the kinetic properties of the full model---such as transition rates and relaxation timescales—requires a more rigorous dynamical framework. The Mori-Zwanzig (MZ) projection formalism~\cite{mori_transport_1965,zwanzig_nonlinear_1973} provides an analytical foundation, offering a systematic way to derive reduced equations that retain essential dynamical information. Several variants of this approach have been developed, including the approach by Gyöngy~\cite{gyongy_mimicking_1986}, the Zwanzig projection as formalized by Legoll and Lelièvre~\cite{legoll_effective_2010,zhang_effective_2016}, as well as hybrid formats~\cite{ayaz_generalized_2022}. For the Zwanzig projection which we consider in this paper, called effective dynamics in~\cite{legoll_effective_2010}, there exist various theoretical results for the error between the original and the CG dynamics~\cite{legoll2019effective,lelievre2019pathwise,legoll2017pathwise,hartmann2020coarse}, see also~\cite{duong_quantification_2018} for  results specific to the Langevin equation and affine CG maps. Importantly, it has been shown that the ability of a CG model to capture rare-event dynamics depends critically on how well the CG map resolves the leading eigenfunctions of the Fokker–Planck or Koopman operator~\cite{zhang_effective_2016, zhang_reliable_2017, nuske_spectral_2021}.

Over the course of the last decade, machine-learning approaches have emerged as powerful tools for constructing dynamically accurate CG models. These include spectral matching~\cite{crommelin_diffusion_2011,nuske_coarse-graining_2019}, generator extended dynamic mode decomposition (gEDMD)~\cite{klus_data-driven_2020}, which is based on the Koopman operator framework~\cite{koopman_hamiltonian_1931, mezic_spectral_2005}, molecular latent space simulators~\cite{sidky_molecular_2020}, implicit transfer operators~\cite{diez2025,diez2025boltzmann,NEURIPS2023_7274ed90}, and reaction coordinate flows~\cite{wu_reaction_2024}. Recently, the gEDMD method was used to directly approximate the coarse-grained dynamics implied by the Zwanzig projection for reversible systems. In the same study, a learning scheme for the parameters of the CG dynamics enabled direct validation of the learned model through rare-event transition timescales~\cite{nateghi_kinetically_2025}. Here, we extend this approach to non-reversible underdamped Langevin dynamics---commonly used to model molecular systems in thermal equilibrium---and integrate it with recent advances in generative modeling, yielding a data-efficient framework for learning coarse-grained Langevin dynamics.

The contributions of this paper are as follows:
\begin{itemize}
    \item We derive, within the Zwanzig projection formalism, explicit analytical expressions for the coarse-grained (CG) dynamics corresponding to the underdamped Langevin equation. The resulting dynamics retain the structure of the full-space system but feature generalized forces and state-dependent diffusion. We further analyze the properties of the associated CG generator.
    \item We show that thermodynamic interpolation (TI)~\cite{moqvist_thermodynamic_2025}, a recently proposed generative modeling approach based on Boltzmann generators~\cite{noe_boltzmann_2019,dibak_temperature_2022, gloy2025hollowflow} and the stochastic interpolant framework~\cite{albergo_stochastic_2023}, can efficiently generate accurate position-space samples for numerically approximating the CG Langevin equation.
    \item We employ the gEDMD method to learn models for the CG Langevin dynamics, and to analyze its dynamical properties, such as rare event transition timescales. Learning algorithms for the CG parameters are provided, and both the gEDMD models and parameters are trained on TI-generated data, removing the need for prohibitively long full-space simulations. 
\end{itemize}

The structure of the paper is as follows: we introduce the  underdamped Langevin dynamics and the key properties of its generator in Section~\ref{sec:langevin_gen}, along with the definition of the phase space CG map which is used throughout this work. The analysis of the CG Langevin dynamics and its parameters is given in Section~\ref{sec:parameters_cg}. In Section~\ref{sec:methods}, we introduce TI as a sampling method, along with the gEDMD algorithm and the learning methods for the CG parameters. Using a two-dimensional model system, we demonstrate the capabilities of the proposed methods numerically, in Section~\ref{sec:numerics}. Detailed proofs for our analytical results, as well as additional simulation and learning details are provided in the Appendix.

\section{Langevin Equation and Coarse Graining Theory}
\label{sec:langevin_gen}
\subsection{Langevin Dynamics}
Underdamped Langevin dynamics are a stochastic version of the Hamiltonian dynamics on full phase space coordinates $(\bq, \bp) \in \R^{2d}$. The dynamical equations are
\begin{equation}
\label{eq:Lan_underdamped}
    \begin{split}
        \diff \bq_t &= \BM^{-1}\bp_t\,\diff t \\
        \diff \bp_t &= -\left[\nabla_\bq V(\bq_t) + \gamma \bp_t \right]\,\diff t +\sqrt{2\gamma\BM\beta^{-1}}\,\diff \BB_t.
    \end{split}
\end{equation}
Here, $\gamma, \beta > 0$ are constants (friction and inverse temperature), $\BM \in \R^{d \times d}$ is a diagonal mass matrix with entries $m_i>0$, while $\BB_t$ is $d$-dimensional Brownian motion. From now on, we will drop the subscript $\bq$ in $\nabla_{\bq} V$ as the potential $V$ is only a function of the position. We apply the same convention to any function that depends only on $\bq$, such as the spatial CG map $\xi$ introduced later. 

We will frequently write~\eqref{eq:Lan_underdamped} in the more compact form 
\begin{equation}
\label{eq:sde}
\diff \BX_t = \bb(\BX_t)\,\diff t + \sqrt{2 \beta^{-1} \BA}\,\diff \BW_t,
\end{equation}
where $\bx = (\bq,\bp) \in \R^{2d}\,, \ \BW_t = (\tilde\BB_t,\BB_t) \in \R^{2d}$ is $2d$-dimensional Brownian motion and the drift and diffusion coefficients are given by 
\begin{align*}
    &\bb(\bq,\bp) = \begin{pmatrix}
        \BM^{-1}\bp \\ - \nabla V(\bq) - \gamma \bp
    \end{pmatrix}  \in \R^{2d}\,, \quad \BA := \sigma\sigma^\top  = \begin{pmatrix}
    0_{d\times d} & 0_{d\times d} \\ 0_{d\times d} & \gamma\BM
\end{pmatrix} \in \R^{2d \times 2d}. 
\end{align*} 
Note that we can also write the drift field $\bb$ in the Hamiltonian form 
\begin{equation*}
    \bb(\bx) = (\BJ - \BA) \nabla H(\bx),   
\end{equation*}
where the canonical Hamiltonian and skew-symmetric matrix $\BJ$ are given by:
\begin{align*}
    H(\bx) = V(\bq) + \frac{1}{2}\bp^T\BM^{-1}\bp \quad \text{ and } \quad \BJ = \begin{pmatrix}
         0_{d\times d} & I_d \\  - I_d & 0_{d\times d} 
\end{pmatrix} \in \R^{2d \times 2d}.
\end{align*}
The invariant distribution of the process~\eqref{eq:Lan_underdamped} is the canonical distribution with density
\begin{equation}
\label{eq:canonical_dist}
\begin{split}
    \diff \mu(\bq, \bp) &\propto \exp\left(-\beta H(\bq,\bp)\right) \diff \bq \diff \bp  \\
    &= \exp\left(-\beta(\frac{1}{2}\bp^T \BM^{-1}\bp + V(\bq))\right)\,\diff \bq\,\diff \bp \\
    &= \mu_\bp(\bp)\mu_\bq(\bq)\,\diff \bq\,\diff \bp.
\end{split}
\end{equation}

\subsection{The Generator}
An important tool for large parts of this paper is the evolution of observables under the Langevin dynamics. Observables are functions $\phi$ of the phase space, which we require to be contained in the weighted space $L^2_\mu(\R^{2d})$:
\begin{equation}
\label{eq:gen_decomp_sym_asym}
    L^2_\mu(\R^{2d}) = \{\phi: \R^{2d} \mapsto \R:\, \vert \, \int \phi^2(\bx) e^{-\beta H(\bx)}\,\diff \bx < \infty \}.
\end{equation}

\noindent The \textbf{generator} is a differential operator acting on observables $\phi \in L^2_\mu(\R^{2d})$:
\begin{equation}     \label{eq:gen_lan}
\begin{split}
    \cL \phi &= \bb \cdot \nabla \phi + \beta^{-1}\BA:\nabla^2 \phi \\ &= \BJ  \nabla H \cdot  \nabla \phi - \BA \nabla H \cdot  \nabla \phi +\beta^{-1}\BA:\nabla^2 \phi,
\end{split}
\end{equation}
where the notation $\BA:\nabla^2 \phi$ denotes the Frobenius inner product between $\BA$ and the Hessian matrix of $\phi$:
\begin{equation*}
    \BA:\nabla^2 \phi = \sum_{i,j=1}^{2d} \BA_{ij} \frac{\partial^2}{\partial \bx_i \partial \bx_j} \phi.
\end{equation*}
The first term in~\eqref{eq:gen_lan} is the Hamiltonian part of the generator 
\begin{equation}
    \cL_{\mathrm{H}} = \BJ \nabla H \cdot \nabla  = \BM^{-1}\bp \cdot \nabla_\bq - \nabla V \cdot \nabla_\bp,
\end{equation}
while the next two terms form the Ornstein-Uhlenbeck part of the generator 
\begin{equation}
    \cL_{\mathrm{OU}} = -\BA \nabla H \cdot \nabla + \beta^{-1} \BA : \nabla^2 = - \gamma \bp \cdot \nabla_\bp + \frac{\gamma}{\beta} \BM :\nabla^2_\bp.
\end{equation}

An equivalent way of decomposing the generator is into its anti-symmetric and symmetric parts:
\begin{equation*}
    \cL = \jstat \cdot \nabla + \frac{1}{\beta}e^{\beta H}\nabla \cdot\left(e^{-\beta H}\BA \nabla \right),
\end{equation*}
where $\jstat$ is the stationary probability flow 
\begin{equation}\label{def:jstat}
    \jstat = \bb - \frac{1}{\beta}e^{\beta H} \nabla \cdot (\BA e^{-\beta H}) = \begin{pmatrix}
        \BM^{-1}\bp \\ - \nabla V(\bq)
    \end{pmatrix}.
\end{equation}
It can be verified directly that $\jstat = \BJ \nabla H$, and therefore $\cL_{\mathrm{H}} = \jstat \cdot \nabla$ coincides with the anti-symmetric part. Similarly, the Ornstein-Uhlenbeck generator $\cL_{\mathrm{OU}}$  coincides with the symmetric part. As a consequence, 
 \begin{equation} \label{eq:prop_full_gen}
\innerprod{\cL \phi}{\psi}_\mu = \innerprod{\jstat \cdot \nabla \phi}{\psi}_\mu - \frac{1}{\beta}\innerprod{\BA\nabla \phi}{\nabla \psi}_\mu, 
\end{equation}
where the rightmost term follows from integration by parts:
\begin{equation*}
\innerprod{\cL_{\mathrm{OU}}\phi}{\psi}_{\mu} = -\frac{\gamma}{\beta}\int_{\R^{2d}} \nabla_\bp \phi \cdot \BM \cdot \nabla_\bp \psi\,\diff \mu(\bq,\bp) 
  =- \frac{1}{\beta}\innerprod{\BA\nabla \phi}{\nabla \psi}_\mu = \innerprod{\phi}{\cL_{\mathrm{OU}}\psi}_{\mu}. 
\end{equation*}
Therefore, in order to approximate the generator, we can use numerical approximations that only require first order derivatives. 

For systems exhibiting meta stable states and undergoing rare transitions, we are generally interested in eigenvalues of the generator close to zero. Thus, we expect to find a number of real eigenvalues $0 = \kappa_0 < \kappa_1 \leq \ldots \leq \kappa_M$ and eigenfunctions $\psi_i, i = 0, \ldots, M$, such that
\begin{equation}
\label{eq:eigenvalue_eq_generator}
    -\cL \psi_i = \kappa_i \psi_i,
\end{equation}
and $\kappa_i \ll 1$. These eigenvalues can be interpreted as rates of transition, their associated implied timescales are defined as~\cite{schutte_direct_1999,prinz_markov_2011}
\begin{equation*}
    t_i = \frac{1}{\kappa_i}.
\end{equation*}

\subsection{CG Map and Potential of Mean Force}\label{ssec:cgmap}
Next, we consider coarse graining of the Langevin dynamics in Equation~\eqref{eq:Lan_underdamped}. We assume the CG map is defined based on a smooth mapping
\[\xi: \R^d \to \R^k\,, \bq \mapsto \xi(\bq), \, k \leq d,\] called the \textbf{spatial CG map}, which transforms the spatial degrees of freedom by mapping them to a lower-dimensional space. In order to arrive at a coarse grained dynamics which is again of Langevin type, we need to define CG velocity. Since formally $ \frac{d}{dt} \xi(\bq) =  \nabla \xi(\bq) \BM^{-1}\bp$, where $\nabla \xi \in \R ^{k\times d}$ is the Jacobian of the spatial CG map, it has been suggested~\cite{duong_quantification_2018,lelievre2012langevin} to define the coarse grained velocity analogously:
\begin{equation}
    \label{eq:cg_momenta}
    \bv_i = \dot{\xi}_i(\bq, \bp) = \nabla \xi_i(\bq) \BM^{-1}\bp,
\end{equation}
leading to an augmented $2k$-dimensional coarse graining map called the \textbf{phase-space CG map}
\begin{equation}
\label{eq:augmented_cg_map}
    \Xi(\bq, \bp) = \begin{bmatrix} \bz \\ \bv \end{bmatrix} = \begin{bmatrix}
        \xi(\bq) \\ \nabla \xi(\bq) \BM^{-1}\bp \end{bmatrix} \in \R^{2k}.
\end{equation}
We use the following notation for its Jacobian determinant, assuming it has maximum rank almost everywhere (i.e.~non-zero determinant for almost all $\bx$):
\begin{equation*}
    \jac(\bx) = \det\left[\nabla \Xi(\bx) \nabla \Xi^\top(\bx) \right].
\end{equation*}

\noindent To proceed further, we need to introduce some abstract concepts. It is often helpful to break up integrals over full phase space into an integral over all points that are assigned to the same point in CG space, and then integrate out over the CG variables. Formally, we write:
\begin{equation*}
\int_{\R^{2d}} g(\bx) \diff \bx = \int_{\R^{2k}} \left[\int_{\R^{2d}} g(\bx) \delta(\by - \Xi(\bx))\, \diff \bx\right] \,\diff \by.
\end{equation*}
To make this notation more transparent, we consider the non-linear pre-image of a point $\by=(\bz,\bv) \in \R^{2k}$ under the CG map as
\begin{equation*}
    \Xi^{-1}(\by) = \left\{\bx \in \R^{2d}: \Xi(\bx) = \by \right\}.    
\end{equation*}
That is, $\Xi^{-1}(\by)$ is the set of phase space elements which collapse to the same point $\by$ in CG space. Writing $\sigma_\by$ for the Hausdorff measure on the nonlinear manifold $\Xi^{-1}(\by)$, we introduce the weighted surface measure:
\begin{equation*}
    \diff \rho_\by(\bx) = \jac^{-1/2}(\bx) \, \diff \sigma_\by(\bx).
\end{equation*}
By means of the measure $\rho_\by$, we now re-write the previous integration formula by splitting up integrals along pre-images of the CG map, known as the \textbf{co-area formula}:
\begin{equation}\label{eq:co_area_formula}
\begin{aligned}
    \int_{\R^{2d}} g(\bx) \diff \bx 
&  = \int_{\R^{2k}} \int_{\Xi^{-1}(\by)} g(\bx) \jac^{-1/2}(\bx)\, \diff \sigma_\by(\bx) \,\diff \by.
\end{aligned}
\end{equation}

\noindent We can now proceed to introduce the \textbf{free energy} or potential of mean force on the CG space:
\begin{equation}\label{eq:free-energy}
    F(\by) = - \frac{1}{\beta}\log \left[\int_{\Xi^{-1}(\by)} e^{-\beta H(\bx)}\,\diff \rho_\by(\bx)\right],
\end{equation}
with associated probability measure $\nu(\by) \sim e^{-\beta F(\by)}\,\diff \by$. The co-area formula implies that for two functions $\phi, \psi \in L^2_\nu(\R^{2k})$, depending only on the CG variables, we have the following identity between inner products on full space and on CG space:
\begin{align*}
    \innerprod{\phi}{\psi}_\nu &= \frac{1}{Z}\int_{\R^{2k}} \phi(\by) \psi(\by) e^{-\beta F(\by)}\,\diff \by\\
    &= \frac{1}{Z}\int_{\R^{2d}} \phi(\Xi(\bx)) \psi(\Xi(\bx)) e^{-\beta H(\bx)}\,\diff \bx = \innerprod{\phi\circ \Xi}{\psi \circ \Xi}_\mu.
\end{align*}

\subsection{Projection Operators and CG Dynamics}
To define dynamics on the CG space, given a phase-space CG map $\Xi$ and the SDE in Equation~\eqref{eq:sde}, the starting point is usually to apply Ito's Lemma to the observed process $\Xi(\BX_t)$:
 \begin{equation*}
     \diff \Xi(\BX_t) = \cL \Xi(\BX_t)\,\diff t + \sqrt{2\beta^{-1}}\nabla \Xi(\BX_t)\cdot \sqrt{\BA}\,\diff \BW_t.
 \end{equation*}
This equation is not closed as it depends on the full state $\BX_t$. A general procedure to close the equation is to apply a projection operator to the infinitesimal generator $\cL$. A linear orthogonal projection $\cP$ is a linear map from functions in $L^2_\mu(\R^{2d})$ to functions in the CG domain $L^2_\nu(\R^{2k})$, which is self-adjoint on $L^2_\mu(\R^{2d})$, and satisfies $\cP^2 = \cP$. A widely used choice for $\cP$ is  \textbf{Zwanzig's projection} operator~\cite{zwanzig_memory_1961}:
\begin{equation}
\label{eq:def_zwanzig_proj}
\begin{split}
    \cP \phi(\by) &= \bE\left[\phi \,\vert\, \Xi(\bx) = \by \right] \\
    &= \frac{1}{e^{-\beta F(\by)}} \int_{\Xi^{-1}(\by)} \phi(\bx) e^{-\beta H(\bx)} \,\diff \rho_\by(\bx).
\end{split}
\end{equation}
The Zwanzig projector represents the best approximation of $\phi$ by a function that only depends on the CG variables $\by$. Following Ref.~\cite{zhang_effective_2016}, we choose $\cP$ according to Eq.~\eqref{eq:def_zwanzig_proj}, and define the generator in CG space by
\begin{equation}
\label{eq:projected_gen_zwanzig}
 \cL^\Xi = \cP \cL \cP.
\end{equation}
It can be shown that $\cL^\Xi$ is indeed the generator of an effective SDE 

\begin{equation}
\label{eq:params_effective_sde}
    \diff \BY_t = \bar\bb(\BY_t) \,\diff t +  \sqrt{2\beta^{-1}}\bar\sigma(\BY_t) \,\diff \BB_t
\end{equation}
on the CG space $\R^{2k}$~\cite{zhang_effective_2016}.

\section{Analytical Structure of the Coarse Grained Langevin Dynamics}
\label{sec:parameters_cg}
Here we report the main analytical results of the paper on the construction of a CG dynamics which preserve structural and dynamical properties of the full-space model.
Following Ref.~\cite{zhang_effective_2016}, the effective drift and diffusion of the coarse grained Langevin equation~\eqref{eq:params_effective_sde} are given by the abstract analytical expressions below:
\begin{align}
    \label{eq:cg-param}
    \bar\bb(\by) &= \cP[\cL \Xi](\by), & \bar \sigma (\bar \sigma)^\top(\by) &= \bar{\BA}(\by) =  \cP[\nabla \Xi \BA \nabla \Xi^\top](\by).
\end{align}
These parameters are spelled out in detail in the following result:
\begin{proposition}
\label{prop:CGdyn}
The effective drift and effective diffusion of the coarse grained Langevin equation can be calculated explicitly as 
\begin{align}
\label{eq:cg_parameters}
    \bar \bb (\bz,\bv)
        &= \begin{pmatrix}
            \bv \\ 
            \bbf(\bz, \bv) - \gamma \bv
        \end{pmatrix}, &
    \bar \BA(\bz,\bv) &= \begin{pmatrix}
        0_{k\times k} & 0_{k\times k} \\ 0_{k\times k} & \gamma \BD(\bz, \bv)
    \end{pmatrix},
\end{align}
where the effective force and diffusion fields are given by
\begin{equation}
\label{eq:force_diff_cg}
\begin{split}
    \bbf(\bz, \bv) &= \cP\left( - \nabla \xi(\bq) \BM^{-1}\nabla V(\bq)  +  \bp^\top \BM^{-1} \partial^2 \xi(\bq) \BM^{-1}\bp \right)(\bz, \bv), \\
    \BD(\bz, \bv) &= \cP\left(\nabla \xi(\bq) \BM^{-1}\nabla \xi(\bq)^\top \right)(\bz, \bv).
\end{split}
\end{equation}
\end{proposition}

In the definition of $\bbf$ we use a compact notation for the Hessian matrix of $\xi$, meaning that the effective force is given component-wise as
\begin{equation*}
    \bbf_i = \cP\left( - \nabla \xi_i \BM^{-1}\nabla V  +  \bp^\top \BM^{-1} \nabla^2 \xi_i \BM^{-1}\bp \right).
\end{equation*}
Proposition \ref{prop:CGdyn} states that the effective SDE~\eqref{eq:params_effective_sde} is again a Langevin dynamics, only with a modified force $\bbf$ and a state-dependent diffusion field $\BD$ acting on the velocity variables. The CG force and diffusion fields in~\eqref{eq:force_diff_cg} essentially capture two effects: first, application of a non-linear change of variables, which manifests in the presence of derivatives of $\xi$. Second, averaging over all phase space points which are collapsed under the action of the CG map $\Xi$, manifest in the action of the projection operator $\cP$.

The structure of the dynamical parameters in Eqs.~(\ref{eq:cg_parameters}-\ref{eq:force_diff_cg}) carries over to the coarse grained generator $\cL^\Xi$ and the invariant distribution of the CG Langevin dynamics:

\begin{proposition}\label{prop:invmeas-gendecomp-CGdyn}
The effective dynamics in Equation~\eqref{eq:params_effective_sde} admits the invariant measure $e^{-\beta F(\by)}$. The associated stationary probability flow is
\begin{equation*}
    \bar{\BJ}^{\mathrm{eq}}(\bz, \bv) = \cP(\jstat \nabla \Xi^\top)(\bz, \bv) = \begin{pmatrix}
            \bv \\
            \bbf(\bz, \bv)
        \end{pmatrix}.
\end{equation*}
Moreover, the projected generator $\cL^{\Xi}$ admits a decomposition:
\begin{align*}
    \cL^{\Xi} = \cL^{\Xi}_a + \cL^{\Xi}_s,
\end{align*}
with anti-symmetric and symmetric parts:
\begin{align}
    \label{eq:gen_cg}
    \cL^{\Xi}_a \phi &= \bar{\BJ}^{\mathrm{eq}} \cdot \nabla_\by \phi, & \cL^{\Xi}_s \phi =& \frac{\gamma}{\beta} e^{\beta F(\by)} \nabla_\bv \cdot \left[ \BD(\by) e^{-\beta F(\by)} \nabla_\bv \phi\right].
\end{align}
\end{proposition}
Proposition~\ref{prop:invmeas-gendecomp-CGdyn} shows that the invariant distribution of the CG dynamics is induced by the potential of mean force, ensuring thermodynamic consistency between the full-space and CG dynamics. As in the full-space setting, the stationary probability flow $\bar{\BJ}^{\mathrm{eq}}$ associated to the invariant distribution determines the anti-symmetric part of the CG generator, as it accounts for the ``Hamiltonian'' part of the CG drift, consisting of the velocity and generalized force terms. The symmetric part is in turn determined by the potential of mean force and the CG diffusion field, and it accounts for the friction term and the generalized diffusion term acting on the CG velocity component. Since $\cL^\Xi_s$ is also given in a divergence form, it admits an integration by parts formula analogous to~\eqref{eq:prop_full_gen}, which can be exploited for numerical approximation of the CG generator.

In summary, Propositions~\ref{prop:CGdyn} and~\ref{prop:invmeas-gendecomp-CGdyn} show that the coarse grained Langevin dynamics retains much of the structure of the full state dynamics~\eqref{eq:Lan_underdamped}. Detailed proofs of these results can be found in Appendix~\ref{sec:proofs}.

\section{Methods}
\label{sec:methods}
We now turn to the numerical approximation and analysis of the coarse grained Langevin equation~\eqref{eq:params_effective_sde}. This entails three essential steps: first, generation of sufficient training data; second, analysis of the coarse grained generator $\cL^\Xi$; third, learning of the parameters $\bar\bb$ and $\bar{\BA}$ of the CG dynamics.

\subsection{Data Generation}
\label{subsec:data_gen}
In order to infer properties of the coarse grained Langevin equation~\eqref{eq:params_effective_sde}, we require samples from the coarse grained equilibrium distribution $\nu(\by)$. As this distribution is the marginal of the full-state canonical distribution $\mu$, it is sufficient to generate data sampling from $\mu$ and subsequently map them by means of $\Xi$. Finally, as the canonical distribution in full space factors into positional component $\mu_\bq$ and momentum component $\mu_\bp$, it is enough to generate samples from $\mu_\bq$ and complement them using samples from the multi-variate normal distribution $\mu_\bp$. To obtain position space samples, we consider three methods:

\subsubsection{Traditional Methods}
\paragraph{Rejection Sampling}
Rejection sampling or acceptance-rejection sampling is a standard method to generate i.i.d. samples from a distribution with known density function (up to normalization constant). On a low-dimensional state space, it is usually an efficient approach, but it is not scalable to higher-dimensional spaces. We use rejection sampling only as a baseline to compare to the performance of other, more scalable methods.

\paragraph{Ergodic Integration of the Langevin Dynamics}
Another standard approach is to integrate the SDE~\eqref{eq:Lan_underdamped} over a sufficiently long time horizon $T$, using an Euler scheme or higher-order integration method. By the ergodic theorem, all snapshots from the trajectory will then form an approximate (correlated) sample from the canonical distribution. As is well known, at high values of $\beta$ (low temperature), a large time horizon $T$ will be needed to explore the entire position space and obtain a near-equilibrated dataset. To circumvent these problems, we employ a generative modeling approach as outlined in the next section.

\subsubsection{Thermodynamic interpolation}

\textbf{Thermodynamic Interpolation} (TI) was proposed in Ref.~\cite{moqvist_thermodynamic_2025} for equilibrium sampling of the Boltzmann distribution across different thermodynamic states. The method is based on flow models where we learn the velocity field $\bb^{(\theta)} : [0,1] \times \mathbb{R}^d \to \mathbb{R}^d$ of an ordinary differential equation (ODE)
\begin{equation}
    \frac{\mathrm{d}\bx(\tau)}{\mathrm{d}\tau} = \bb^{(\theta)}\left(\tau,\bx(\tau)\right), \bx(0) = \bx_0,\label{eq:cnf}
\end{equation}
 such that its solution is a flow map $\Phi_{\tau}^{(\theta)}$ with a push-forward $\Phi_{1\,\#}^{(\theta)} \rho_0 = \rho_1$. In other words, if we draw a sample $\bx_0 \sim \rho_0$, and integrate Eq.~\eqref{eq:cnf} for $\tau$ from $0$ to $1$, we obtain a sample from $\bx_1 \sim \rho_1$. In Ref.~\cite{moqvist_thermodynamic_2025}, two versions of latent and ambient TI were proposed. Here, we apply the ambient version, meaning that the initial distribution is the canonical position space distribution $\rho_0 = \mu_\bq(\beta_0)$ at a high temperature $\beta_0$, while the target is the canonical distribution at a low temperature, i.e. $\rho_1 = \mu_\bq(\beta_1)$. We choose the ambient version over the latent one because it extrapolates better to $\beta$ values outside the training range and exhibits superior empirical performance.

Our implementation of TI is based on the two-sided interpolant~\cite{albergo_stochastic_2023}
\begin{equation}\label{eq:interpolant_def}
    \bx(\tau) = I(\tau, \bx_0, \bx_1) + \gamma(\tau) \bz
\end{equation}
where $\bz \sim \mathcal{N}(0,\text{Id})$, $\bx_0 \sim \rho_0$, $\bx_1 \sim \rho_1$. The function $I$ is a deterministic interpolant that satisfies $\bx(0)=\bx_0$, $\bx(1)=\bx_1$, and $\gamma(\tau)$ is a smooth function controlling the noise level at intermediate times (see~\cite{albergo_stochastic_2023} for details.) We train the model according to the Stochastic Interpolant loss function

\begin{equation}
\begin{split}
    & \BL\left[\bb^{(\theta)} \right]\\
    &= \bE_{\tau,\bz,\bx_0,\bx_1} \left[\frac{1}{2}|\bb^{(\theta)}(\tau, \bx(\tau); \beta_0, \beta_1)|^2 - \bb^{(\theta)}(\tau, \bx(\tau); \beta_0, \beta_1) \cdot \left( \partial_\tau I(\tau,\bx_0, \bx_1)+\dot{\gamma}(\tau)\bz \right)\right],
\end{split}
\end{equation}
employing antithetic sampling for smoother convergence, as previously described \cite{albergo_stochastic_2023}. The loss function essentially minimizes the error between the learned velocity field and true one based on the interpolant. The expectation in the loss is taken independently over time $\tau$, the random variable $z$, $x_0$ drawn from initial distribution, and $x_1$ drawn from the target distribution.

Once the TI model has been trained, we can generate samples from the target distribution (lower temperature) given initial states from the initial distribution (higher temperature). We also compute the \textit{importance weight} of each sample as proposed in Ref.~\cite{moqvist_thermodynamic_2025}, in order to account for any bias or inaccuracy in the model output distribution. These weights are later used to correct towards the equilibrium expectation of mass and stiffness matrices $\BG$ and $\BA$ within the gEDMD algorithm (see next section), in the same fashion as in Ref.~\cite{noe_boltzmann_2019}. It is worth noting that the weights are computed with respect to the unnormalized distribution, as the normalization constant cancels out in the gEDMD algorithm.

\subsection{gEDMD}
In order to verify if meta stable states, eigenvalues and implied timescales are preserved by the CG dynamics, we need to numerically approximate the coarse grained infinitesimal generator $\cL^\Xi$ in Equation~\eqref{eq:projected_gen_zwanzig}. As shown in Ref.~\cite{klus_data-driven_2020}, $\cL^\Xi$ can be approximated by a data-driven method called generator Extended Dynamic Mode Decomposition (gEDMD). Remarkably, the gEDMD algorithm can be applied directly on the full-space data without the need to compute the coefficients $\bar\bb$ and $\bar{\BA}$ of the CG dynamics, see also Ref.~\cite{nateghi_kinetically_2025}. 

The algorithm requires $n$ basis functions acting on the CG phase-space variables $\bz,\, \bv$:
\begin{equation*}
    \phi(\bz,\bv) = \{\phi_1(\bz,\bv), ..., \phi_n(\bz,\bv)\},    
\end{equation*}
along with position space samples $\{\bq_l\}_{l=1}^m$, and the Jacobians of the CG map $\nabla \xi(\bq_l)$ evaluated at each sample. We then augment these data by drawing momenta from the canonical momentum distribution, yielding phase space samples $\bx_l = (\bq_l, \bp_l)$. We  project these samples into CG phase space as 
\begin{equation*}
\by_l = 
    \begin{pmatrix}
        \bz_l \\ \bv_l
    \end{pmatrix} =
    \begin{pmatrix}
        \xi(\bq_l) \\ \nabla \xi(\bq_l) \BM^{-1}\bp_l
    \end{pmatrix}.
\end{equation*}
Next, we build $n\times m$-dimensional data matrices
 \begin{align*}
     \mathbf{\Phi} &= \begin{bmatrix}
         \phi(\bz_1, \bv_1) & \vert & \cdots &\vert& \phi(\bz_m, \bv_m)
     \end{bmatrix} &
     \cL\mathbf{\Phi} &= \begin{bmatrix}
         [\cL\phi](\bq_1, \bp_1) & \vert & \cdots &\vert& [\cL\phi](\bq_m, \bp_m)
     \end{bmatrix}.
 \end{align*}
To form the matrix $\cL\mathbf{\Phi}$, we apply the full-state generator $\cL$, which requires derivatives of the spatial CG map. Note that when we apply the full-state generator to functions acting on the CG space, we treat them as the composition $\phi \circ \Xi$. The final matrix approximation $\BL$ to the CG generator $\cL^\Xi$ is then given by
\begin{equation}
\label{eq:gen-ag}
 \widehat{\mathbf{L}} = \widehat{\BG}^{-1}\widehat{\BA},
\end{equation}
 where
 \begin{align}
    \widehat{\BG} &= \frac{1}{m}\sum_{l=1}^m \phi(\by_l) \otimes \phi(\by_l), & \widehat{\BA} &= \frac{1}{m}\sum_{l=1}^m \phi(\by_l) \otimes [\cL\phi](\bx_l)
    \label{eq:gen-AG-est}
\end{align}
are called mass and stiffness matrices, respectively. To solve for $\widehat{\BL}$ numerically, we do not explicitly form $\widehat{\BG}$, instead, we perform a whitening transformation by removing the small singular values of $\mathbf{\Phi}$:
\begin{align}
    \mathbf{\Phi} &= \BU \BS \BV^\top, &
    \BR &= \BU\BS^{-1} \in  \mathbb{R}^{n\times r}, &
    \widehat{\BL}_r &= \BR^\top \widehat{\BA}\hspace{0.1cm} \BR,
    \label{eq:gen-red}
\end{align}
where $r\leq n$. In practice, we do not fix $r$ in advance and instead set a threshold to cut off small singular values. Slowest timescales $t_i$ of the system, or dominant eigenvalues $\kappa_i$ of the generator, can be computed by diagonalization of the reduced matrix $\widehat{\BL}_r$.

In this study, we use random Fourier features (RFFs)~\cite{rahimi_random_2007} as basis functions: 
\begin{equation*}
    \phi(\by) = \{\cos{(\omega^{\top}_1 \by)}, \sin{(\omega^{\top}_1 \by)}, \dots, \cos{(\omega^{\top}_n \by)}, \sin{(\omega^{\top}_n \by)} \},   
\end{equation*}
where $\{\omega_1, \dots, \omega_n\}$ are random frequency vectors drawn from a spectral distribution $\rho$. RFFs represent a low-rank approximation to a reproducing kernel function, e.g. associated to a Gaussian kernel~\cite{rahimi_random_2007}. The number of basis functions $n$, or number of random frequencies in the case of RFFs, and any parameter associated to the underlying kernel are hyperparameters that need to be optimized.

\subsection{Parameter Estimation for CG Dynamics}
\label{subsec:lea_meth}
To produce simulations of the coarse grained Langevin SDE, we need to learn models for their diffusion and drift coefficients defined in Equation~\eqref{eq:cg_parameters}. Let us define the \textit{local force} $\bbf_{\mathrm{loc}}$ and \textit{local diffusion} $\BD_{\mathrm{loc}}$ as follows:
\begin{equation}
\label{eq:force_diff_local}
\begin{split}
    \bbf_{\mathrm{loc}}(\bq, \bp) &= - \nabla \xi(\bq) \BM^{-1}\nabla V(\bq)  +  \bp^\top \BM^{-1} \partial^2 \xi(\bq) \BM^{-1} \bp, \\
    \BD_{\mathrm{loc}}(\bq, \bp) &= \nabla \xi(\bq) \BM^{-1}\nabla \xi(\bq)^\top.
\end{split}
\end{equation}
These are the terms which are projected onto CG space in Eq.~\eqref{eq:force_diff_cg} to form the learnable parts of the effective drift (velocity component) and diffusion (lower right block). Using available training data, we can learn the effective drift and diffusion as best approximations of the local force and diffusion, by solving data-driven regression problems, similar to the one proposed in Ref.~\cite{nateghi_kinetically_2025}:
\begin{equation}
    \label{eq:minimization_diff}
\begin{split}
    \BD_m & = \underset{\tilde{\BD} = \tilde{\BD}(\bz,\bv)}{\argmin} \frac{1}{m}\sum^m_{l=1} \left\|\tilde{\BD}(\bz_l, \bv_l) - \BD_{\mathrm{loc}}(\bq_l,\bp_l) \right\|_F ^ 2 \\
    \bbf_m & = \underset{\tilde{\bbf} = \tilde{\bbf}(\bz,\bv)}{\argmin} \frac{1}{m}\sum^m_{l=1} \left\|\tilde{\bbf}(\bz_l, \bv_l) - \bbf_{\mathrm{loc}}(\bq_l,\bp_l) \right\|_F ^ 2,
\end{split}
\end{equation}
where the regression is carried out against $m$ samples. Note that the learned effective force and diffusion are only functions of the CG variables. The complete effective drift and diffusion fields can then be constructed as
\begin{align*}
        \bar{\bb}_m &= \begin{pmatrix}
            \bv \\ \bbf_m(\bz, \bv) - \gamma \bv
        \end{pmatrix}, &
        \bar{\BA}_m &= \begin{pmatrix}
            0_{k \times k} & 0_{k \times k} \\
            0_{k \times k} & \gamma\BD_m(\bz, \bv)
        \end{pmatrix}.
\end{align*}

In the numerical examples, we parametrize the effective force and diffusion either as a linear combination of random Fourier features, or as the output of a shallow neural network (NN). We found that the solution of the RFF regression tends to be accurate within data-rich regions, while the NN provides better extrapolation accuracy at the cost of increased computational cost. During simulations of the CG dynamics, we apply a hybrid method to leverage the strengths of both models by a switching rule: the RFF-based solution is used within data-rich regions, while the more accurate NN is used for extrapolation outside this region.

The summary of the numerical approximation of the CG dynamics is given in Algorithm~\ref{alg:gEDMD_1}.

\begin{figure}
\begin{algorithm}[H]
\caption{Numerical Approximation of the Coarse Grained Langevin Equation\label{alg:gEDMD_1}}
\begin{tabular}{ll}
    \textbf{Input:} & position space samples $\{\bq_l\}_{l=1}^m$,
    \,basis functions $\phi(\bz, \bv) \in \R^n$.
  \end{tabular}
\begin{algorithmic}[1]
\State \textbf{Data Augmentation:}
\State Complement position space samples by random momenta: $\bx_l = (\bq_l, \bp_l)$.
\State Convert augmented samples into CG phase space samples:
$\by_l = (\bz_l, \bv_l) = \Xi(\bq_l, \bp_l)$.
\State \textbf{Generator Learning (gEDMD):}
\State Compute mass and stiffness matrices for the generator:
\begin{align*}
\hat{\BG} &= \frac{1}{m}\sum_{l=1}^m \phi(\by_l) \otimes \phi(\by_l),
& \hat{\BA} &= \frac{1}{m}\sum_{l=1}^m \phi(\by_l) \otimes [\cL\phi](\bx_l).
\end{align*}
\State Compute generator matrix $\hat{\BL} = \hat{\BG}^{-1}\hat{\BA}$.
\State \textbf{Parameter Estimation:}
\State Learn a parametric model for the effective force $(\bar{\bbf}_m)$ and diffusion $(\bar{\BA}_m)$, as in Equation~\eqref{eq:minimization_diff}.
\end{algorithmic}
\end{algorithm}
\end{figure}

\section{Numerical Example}
\label{sec:numerics}
We illustrate the coarse graining scheme discussed in Sections~\ref{sec:langevin_gen} and \ref{sec:parameters_cg} and the proposed numerical methods in Section~\ref{sec:methods} using the two-dimensional Lemon Slice potential.

\subsection{Lemon Slice potential}
\begin{figure}[ht]
    \centering
    \includegraphics[width=0.5\linewidth]{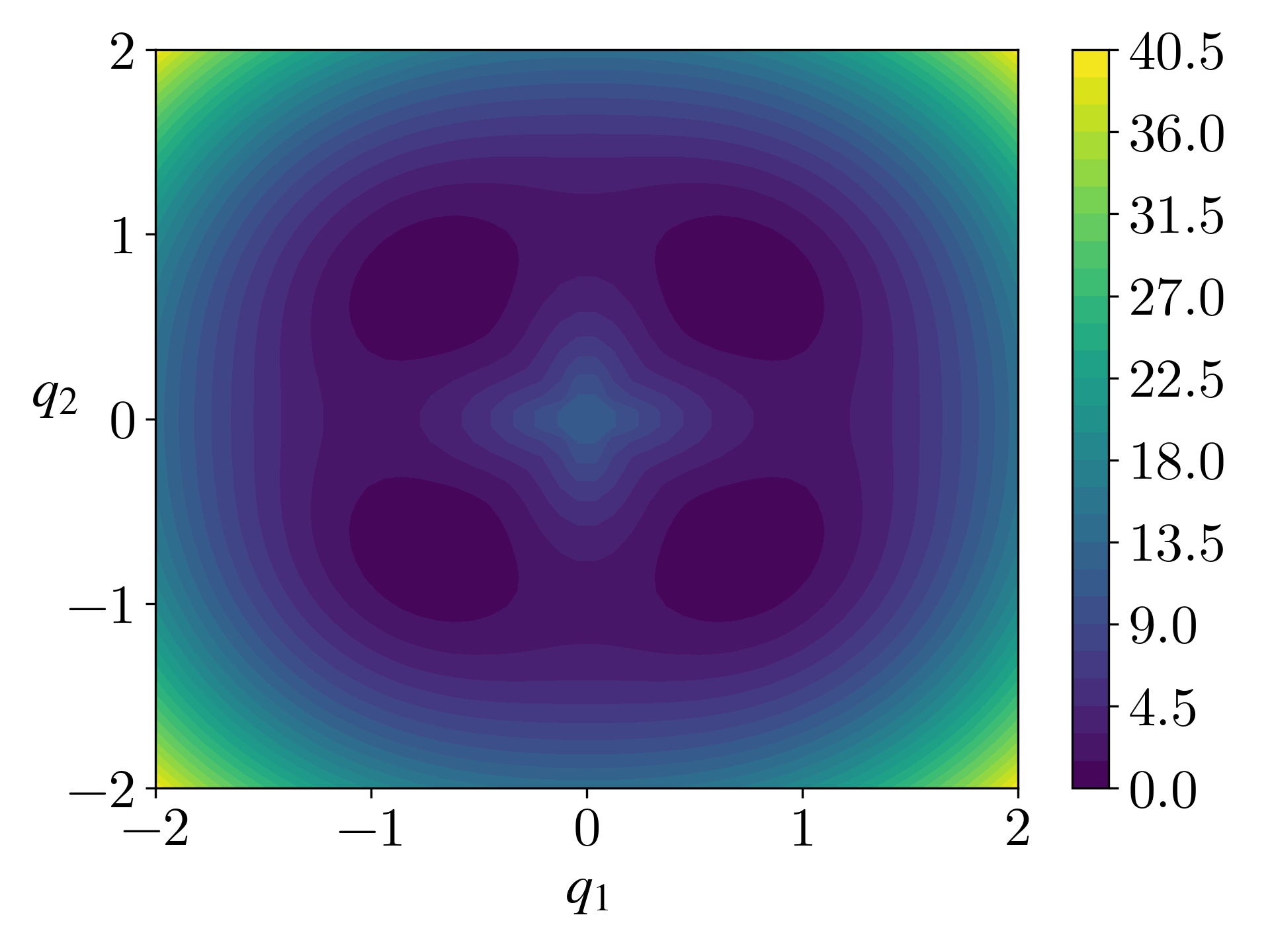}
    \caption{Potential field of the Lemon Slice system. }
    \label{fig:lemon-system}
\end{figure}

The Lemon Slice potential field is a function of two-dimensional polar coordinates $r$ and $\theta$ by
\begin{equation}
	V(r,\theta) = 1.2 \cdot [\cos(4\theta) + 10(r-1)^2].
\end{equation}
 The energy landscape of the system is shown in Figure~\ref{fig:lemon-system}, consisting of $4$ minima, which correspond to meta stable states at low temperature.

The potential field gives rise to underdamped Langevin dynamics~\eqref{eq:Lan_underdamped} on four-dimensional phase space $(\bq,\bp) \in \R^4$.

\begin{figure}[ht]
    \centering
    \includegraphics[width=0.6\linewidth]{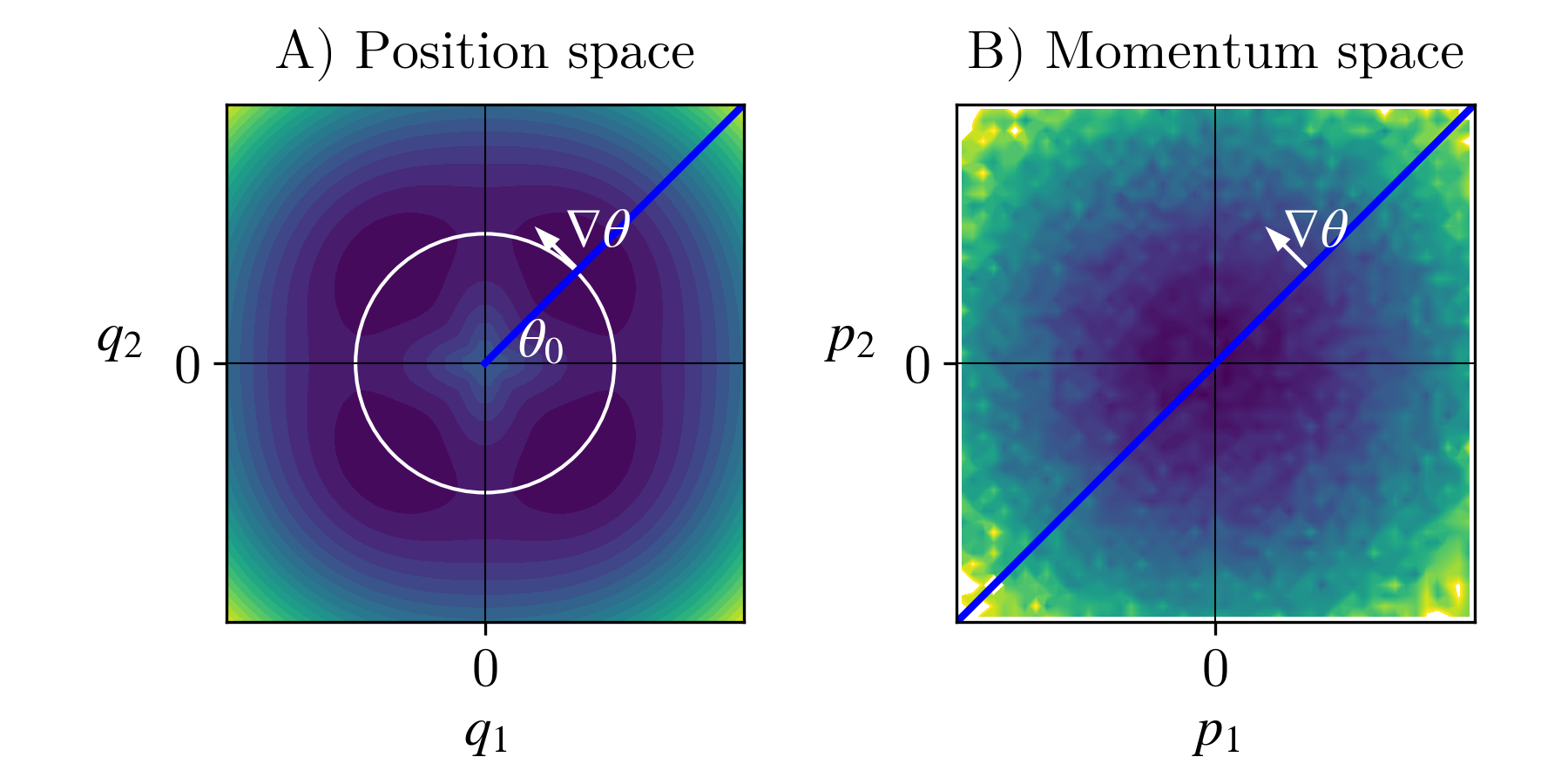}
    \caption{CG definition for the Lemon Slice system, in position and momentum space. Points belonging to the blue line in positional space have spatial CG component of $\xi(r, \theta) = \theta_0$, and points belonging to the blue line in the momentum space are perpendicular to $\nabla\theta$, and therefore have zero velocity CG component. Note that for $\bv \neq 0$, the pre-image in momentum space would be more complex as the magnitude of the gradient $\nabla \theta$ depends on $r$.}
    \label{fig:lemon-system-cg}
\end{figure}

Following previous work~\cite{nuske_spectral_2021,nateghi_kinetically_2025}, the polar angle $\theta$ is a suitable spatial CG coordinate for this system as it separates the four minima. According to Equation~\eqref{eq:cg_momenta}, we can define the phase-space CG map as
\begin{align*}
    \Xi(\bq,\bp) &= \begin{pmatrix}
        \xi(\bq) \\ \nabla \xi(\bq) \BM^{-1}\bp
    \end{pmatrix}, & \xi(\bq)=\theta=\mathrm{atan}\left(\frac{\bq_2}{\bq_1}\right).
\end{align*}
Figure~\ref{fig:lemon-system-cg} shows an illustration of the phase-space CG map for $\BM = \mathrm{Id}$. For a fixed point in CG space at $(\bz,\bv)=(\theta_0,0)$, the level set $\Xi^{-1}(\theta_0, 0)$ corresponds to the Cartesian product of the blue lines shown in position and momentum space, respectively.
We prepare training datasets based on all methods reported in Section~\ref{subsec:data_gen}: rejection sampling, ergodic simulations, and thermodynamic interpolation (see next section for more details). The simulation details, as well as the selected parameters are given in Appendix~\ref{app:sim_set}.

\subsection{Timescale Analysis}
We begin by comparing the implied timescales of the full four-dimensional Langevin dynamics~\eqref{eq:Lan_underdamped} and the coarse grained dynamics~\eqref{eq:params_effective_sde}. We also study their dependence on the inverse temperature $\beta$.

\begin{figure}[ht]
    \centering
    \includegraphics[width=0.7\linewidth]{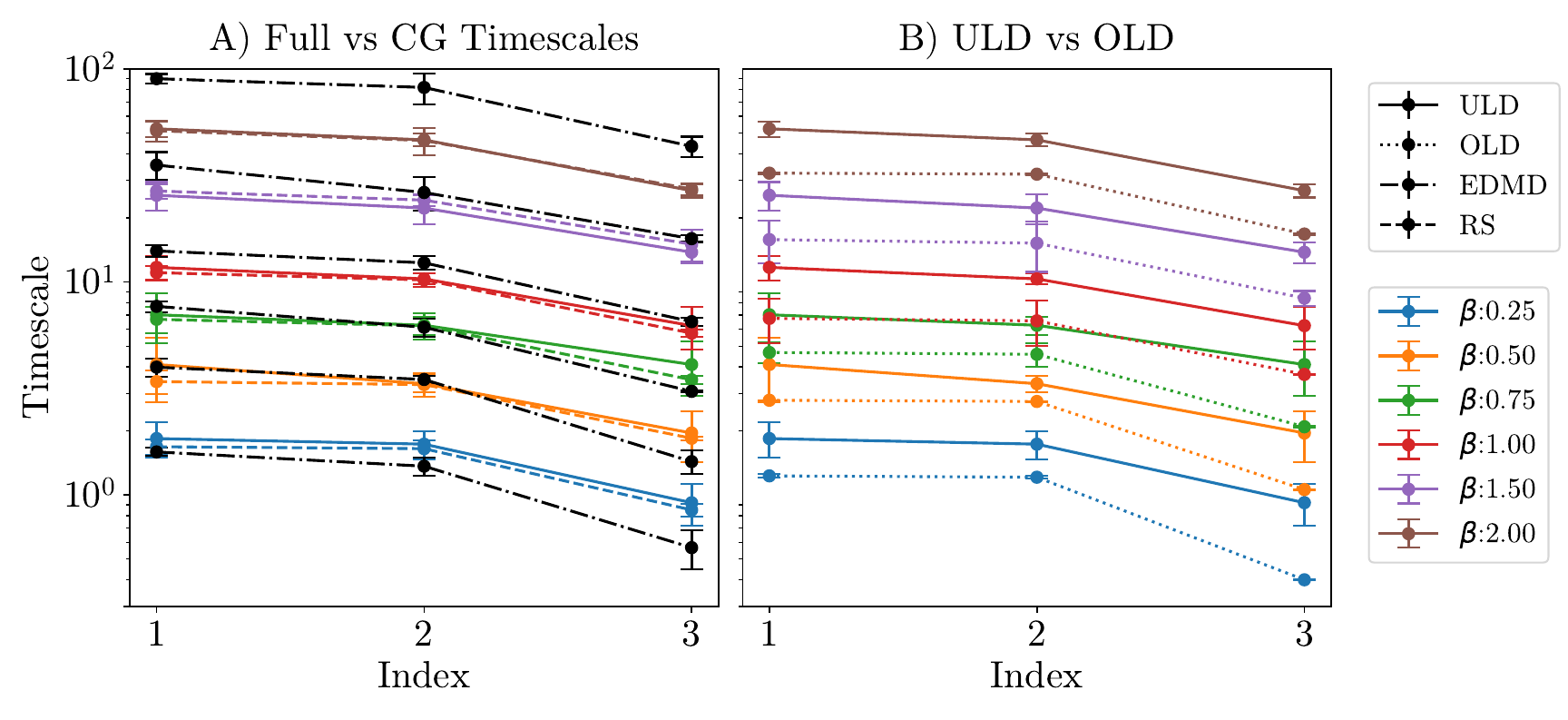}
    \caption{\textbf{A)} Slowest timescales $t_1$ to $t_3$ of the coarse grained Langevin equation~\eqref{eq:params_effective_sde} with the Lemon slice potential, corresponding to different values of the inverse temperature $\beta$, computed on rejection sampling data (RS, dashed lines) or on simulations of the underdamped Langevin SDE (ULD, solid lines), compared against the baseline timescales from the EDMD method (EDMD, black dash-dotted lines). \textbf{B)} timescales of the underdamped CG Langevin dynamics (ULD, solid lines) versus overdamped CG Langevin dynamics (OLD, dotted lines).}
    \label{fig:its-comp}
\end{figure}

\paragraph{Identical Masses} We first study the case where $\BM$ is the identity matrix. To obtain a baseline, we apply the EDMD method~\cite{williams_data-driven_2015} to the long ergodic simulation data, and compute the eigenvalues and timescales of the full system's Koopman operator. The basis set for EDMD consists of random Fourier features (RFF) on four-dimensional phase space, which have been shown to provide accurate approximations of eigenvalues and timescales~\cite{nuske_efficient_2023}. The dominant timescales for inverse temperature values in a range $\beta \in [0.25, 2.0]$ are shown in Figure~\ref{fig:its-comp}. The results show that for the selected inverse temperature range, the timescales vary by about two orders of magnitude, which proves this setting to be an interesting test case for the effectiveness of sampling algorithms.

Next, we compute an approximation of the projected generator $\cL^\Xi$ following Algorithm~\ref{alg:gEDMD_1}, using both datasets obtained by rejection sampling and by ergodic simulation of the full Langevin dynamics. Again we use RFF basis functions, their main hyper-parameters (the number of random features and the kernel bandwidth) are optimized using cross validation based on the VAMP-score. Details of this analysis are given in Appendix~\ref{app:vamp}. We then solve the associated eigenvalue problem for the projected generator, which provides the eigenvalues and implied transition timescales of the CG dynamics~\eqref{eq:params_effective_sde}. The results depicted in Figure~\ref{fig:its-comp} show agreement of the CG generator's timescales with the baseline, regardless of the data generation scheme. We conclude that, if the spatial CG map is chosen well, the underdamped CG dynamics~\eqref{eq:params_effective_sde} can be expected to retain slow transition timescales of the full system. Moreover, we can use the gEDMD algorithm to analyze the properties of the CG Langevin dynamics.

We also compute the timescales of the system if it was governed by overdamped Langevin dynamics in position space only. The results show a similar temperature dependence, and are almost identical but slightly smaller compared to the underdamped case. In previous work, it was observed that overdamped dynamics lead to an acceleration of the timescales~\cite{nateghi_kinetically_2025}. The effect is not drastic for this model system, but can be expected to be much more noticeable for larger-scale systems. \\

\begin{figure}[ht]
    \centering
    \begin{subfigure}[c]{0.35\textwidth}
    \includegraphics[width=\linewidth]{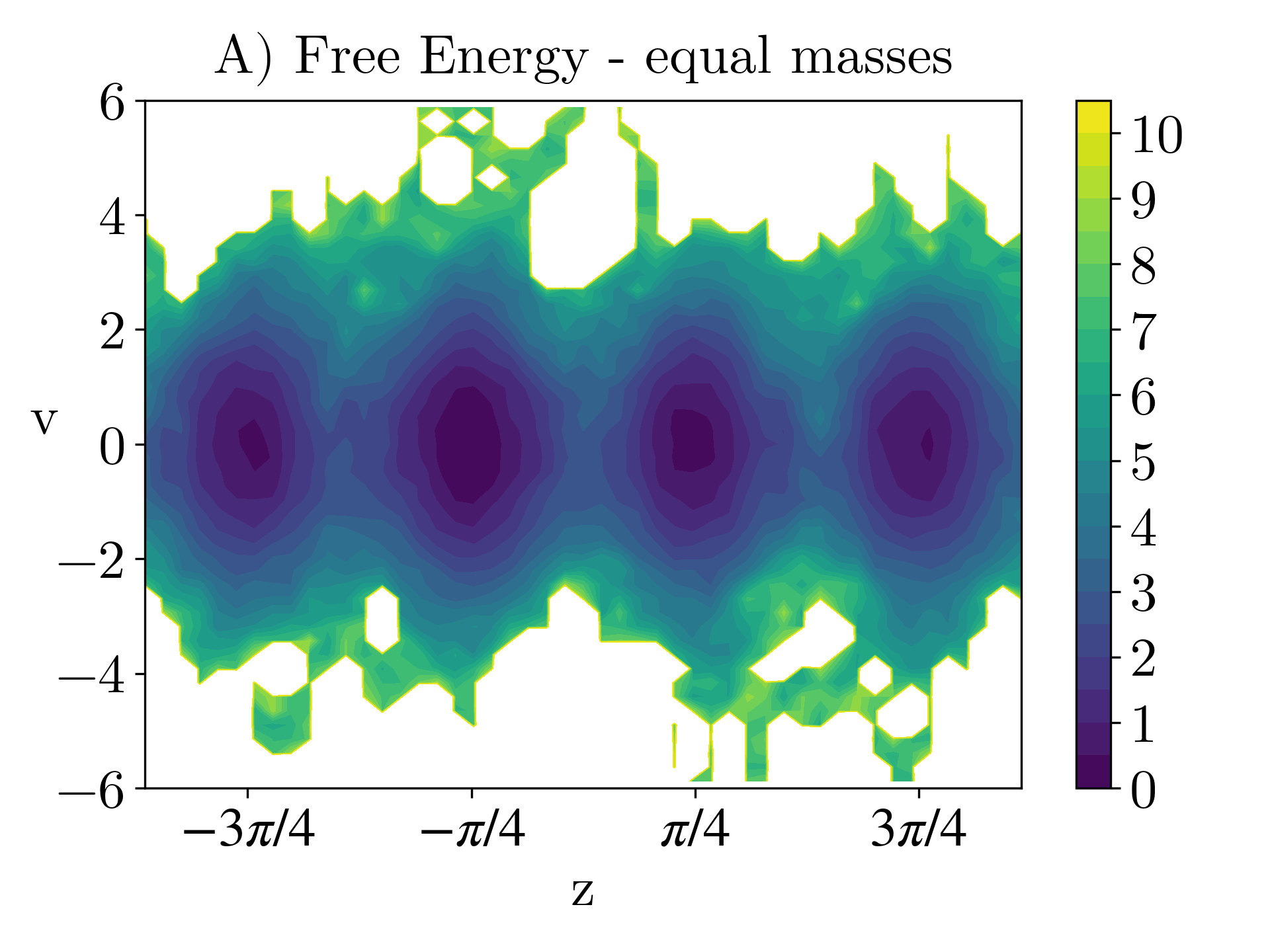}
    \end{subfigure}
    \begin{subfigure}[c]{0.35\textwidth}
    \includegraphics[width=\linewidth]{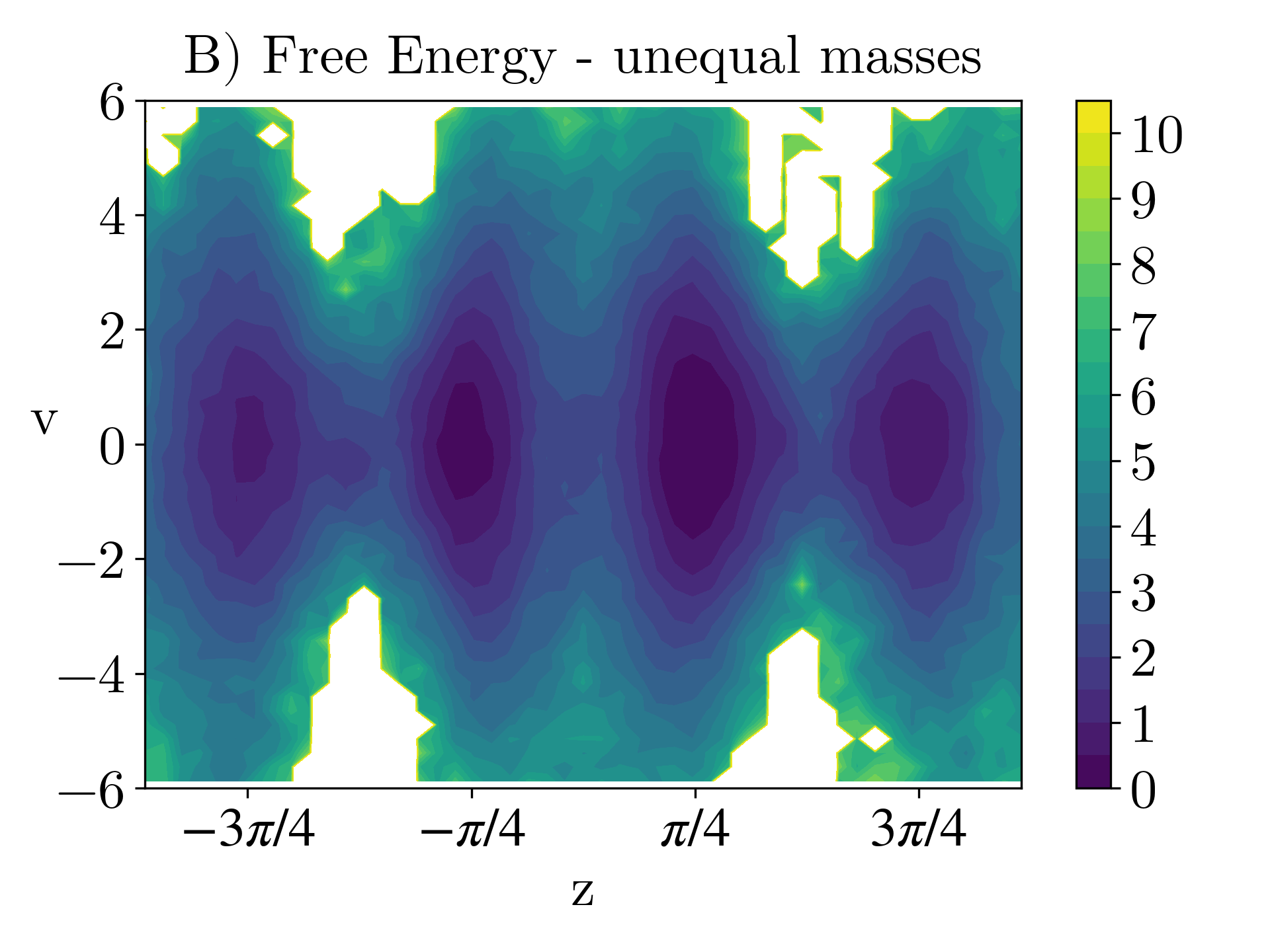}
    \end{subfigure}
    \begin{subfigure}[c]{0.35\textwidth}
    \includegraphics[width=\linewidth]{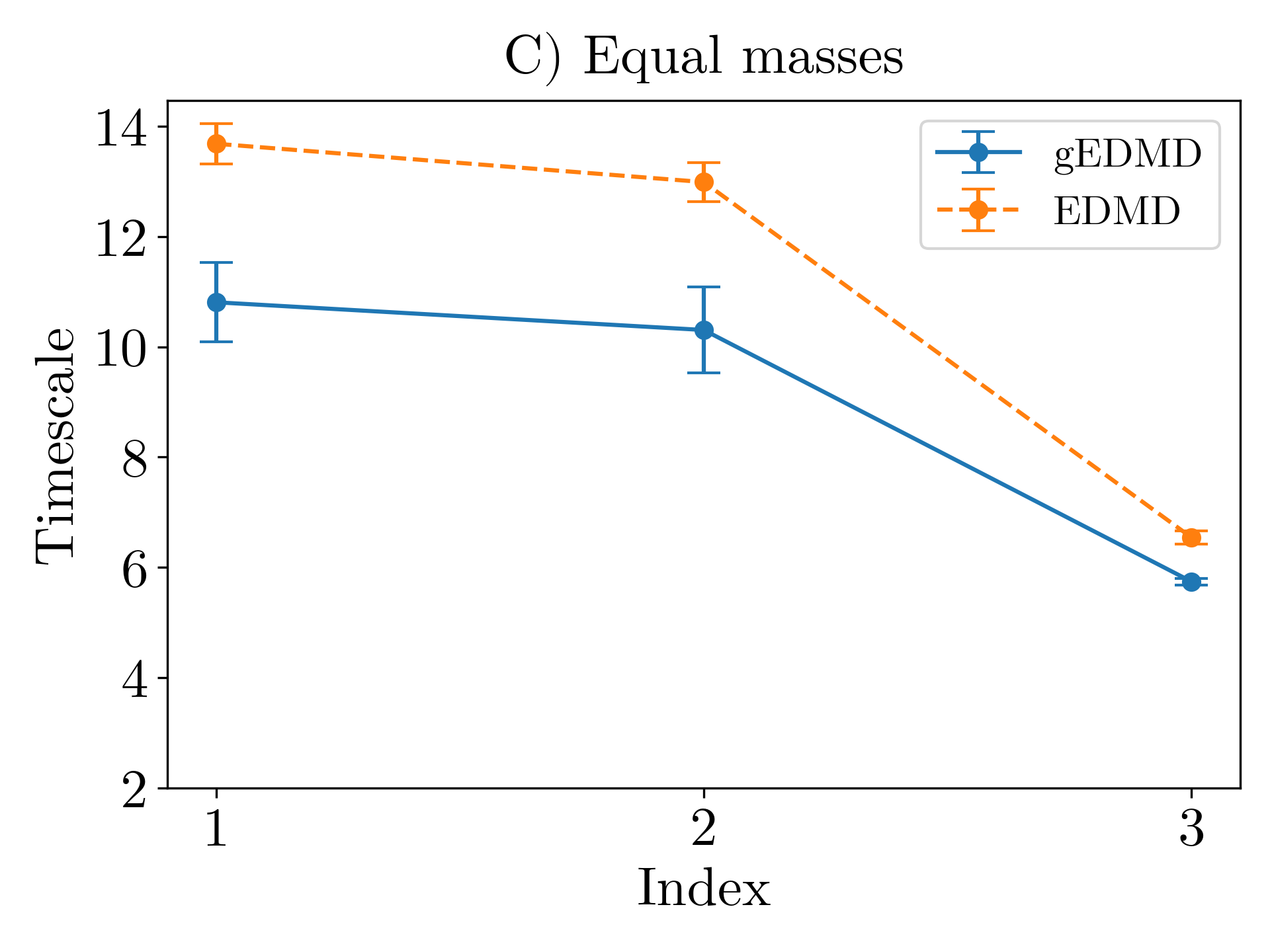}
    \end{subfigure}
    \begin{subfigure}[c]{0.35\textwidth}
    \includegraphics[width=\linewidth]{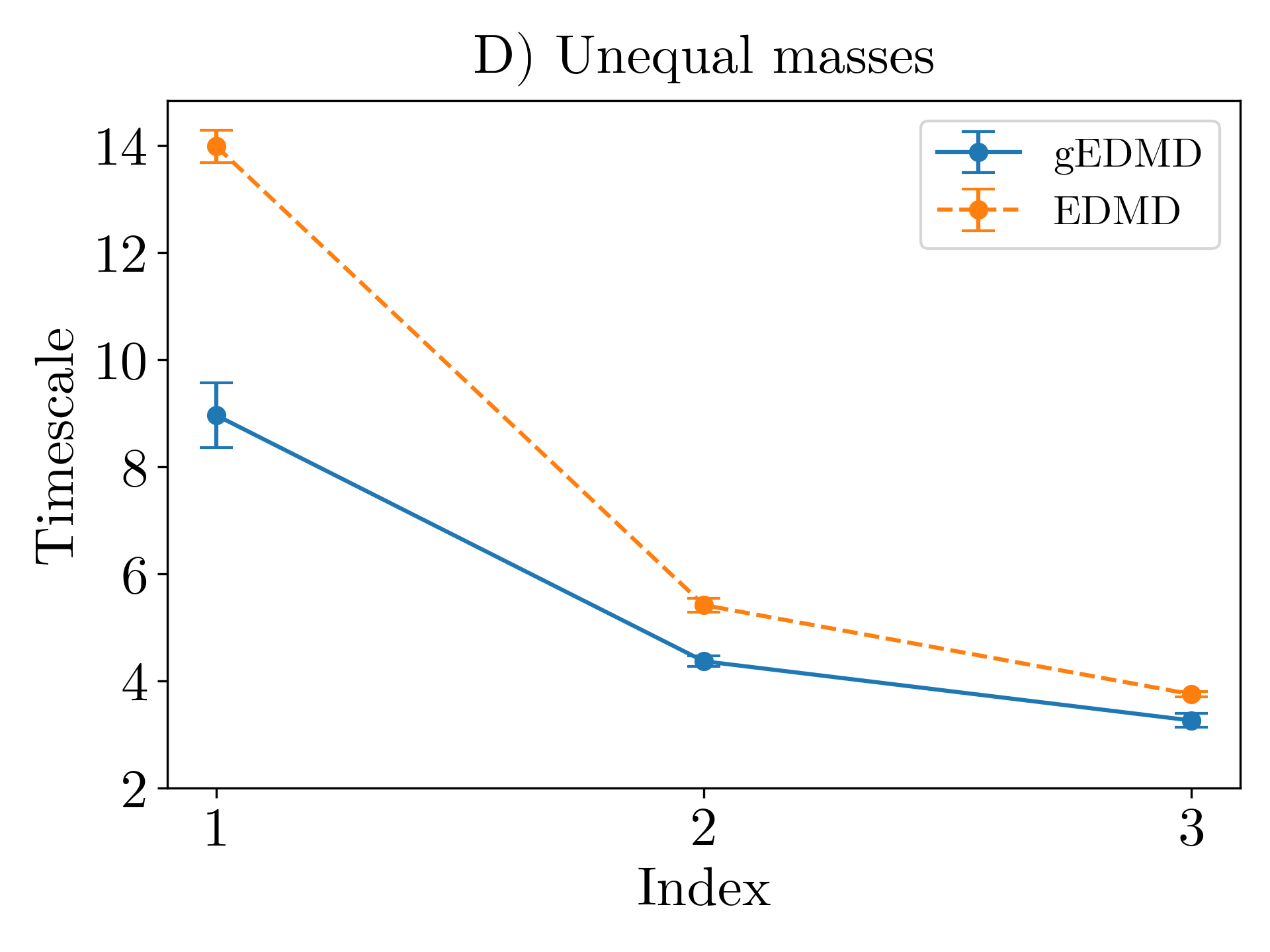}
    \end{subfigure}
    \caption{\textbf{Top Row:} Empirical free energy surface for lemon slice with identity mass matrix in (\textbf{A}) and with unequal masses in (\textbf{B}), both at $\beta=1$. \textbf{Bottom Row}: Slowest timescales $t_1$ to $t_3$ of the coarse grained Lemon slice system with equal masses in (\textbf{C}) and unequal masses in (\textbf{D}), both based on the integration of the underdamped Langevin SDE at $\beta=1$.}
    \label{fig:diffmass_fes}
\end{figure}

\paragraph{Unequal Masses} Next, we study the case where different masses act along different coordinate directions by setting $\BM=\text{diag}(1.00, 0.25)$, meaning that the mass in the horizontal direction is larger than along the vertical direction. The free energy surface in CG space is depicted in Figures~\ref{fig:diffmass_fes}A and~\ref{fig:diffmass_fes}B. The full space momentum distribution is more spread out in the $\bp_2$-direction for decreased $m_2$, and hence we expect to see a larger spread of the $\bv$-distribution in CG space, which is indeed observed in the free energy surface.

Next, we again approximate the projected generator $\cL^{\Xi}$ to obtain the timescales of this system. Figure~\ref{fig:diffmass_fes}D show the dominant timescales computed via the gEDMD algorithm compared to EDMD timescales serving as a reference. Both sets of timescales are once again in agreement. With the introduction of unequal masses, the first and second non-trivial timescales are no longer identical, as the unequal masses break the symmetry of the system. As a result, the transition from the left- to the right-hand side of the potential field is slower than the transition from the bottom to top part of the potential field, leading to a down-shift of $t_2$.

\subsection{Application of Thermodynamic Interpolation}
Next, we show that thermodynamic interpolation can produce accurate position space samples outside its training regime, allowing to estimate transition timescales using gEDMD.  Returning to the case where $\BM = \mathrm{Id}$, the TI model is trained against a dataset consisting of samples from a simulation of the Lemon Slice system at high temperatures $\beta \in \left[0.25, 0.50, 0.75, 1.00\right].$ In each training step, we choose two specific values of $\beta$ from this set, along with one sample corresponding to each of these values, and ask the model to interpolate between them. This process is repeated while shuffling through different combinations of $\beta$. The model is then validated against simulation data corresponding to low temperature $\beta = 2.00$ interpolated from $\beta=1.00$. Further details on the TI architecture and training are provided in Appendix~\ref{app:ti}. 

\begin{figure}[ht]
    \centering
    \begin{subfigure}[c]{0.6\textwidth}
    \includegraphics[width=\linewidth]{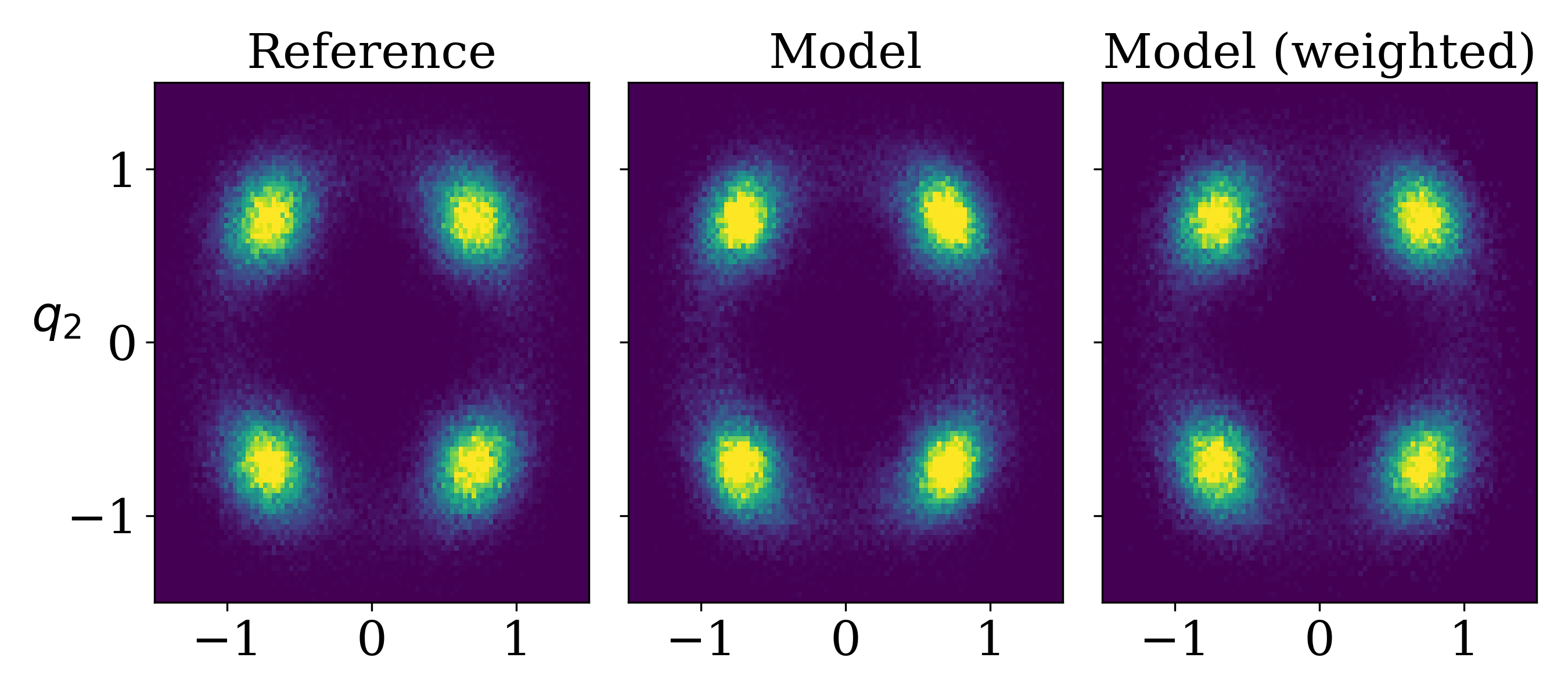}   
    \end{subfigure}
        \begin{subfigure}[c]{0.6\textwidth}
        \includegraphics[width=\linewidth]{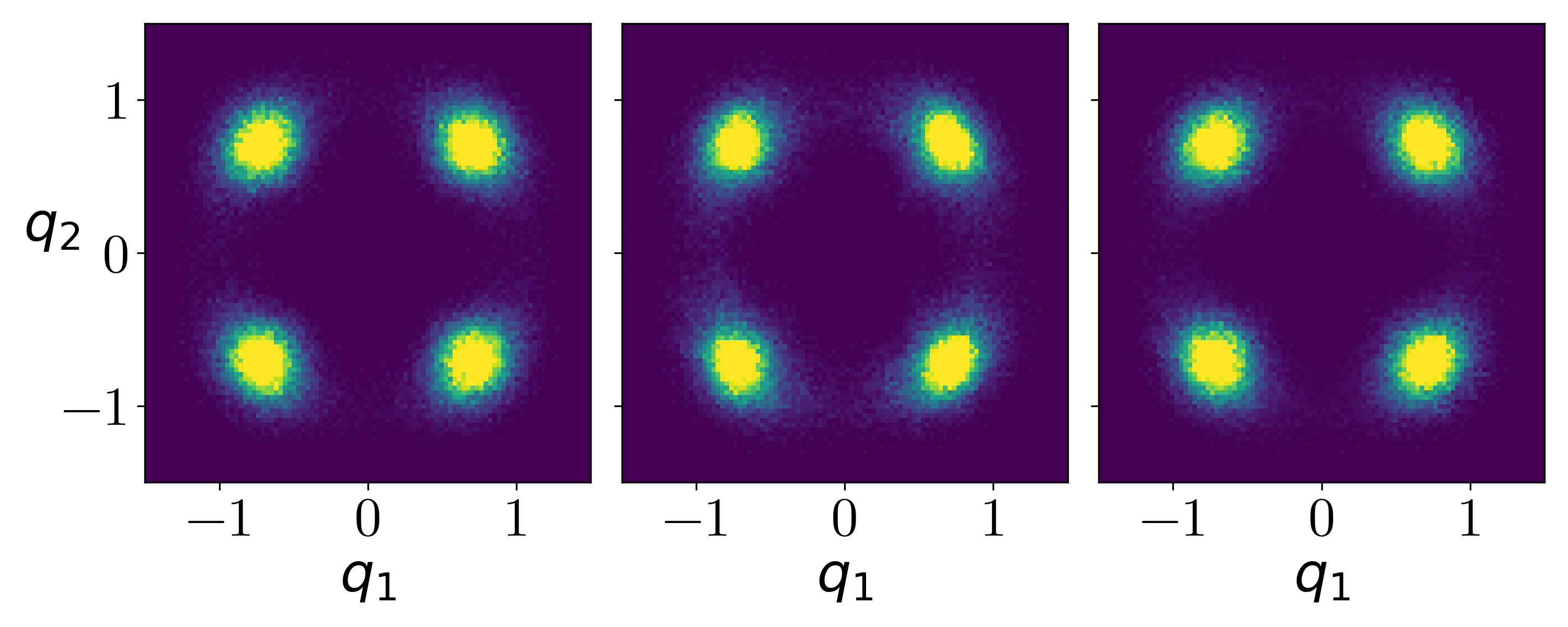}
        \end{subfigure}
    \caption{Histograms comparing the dataset obtained via rejection sampling (left) with data generated by the TI model without (middle) and with (right) importance sampling weights. First row is for the case of $\beta=1.50$ and the second row is for the case of $\beta=2.00$.}
    \label{fig:genai_histo}
\end{figure}

Once we have trained the generative model, we use it to produce samples at various temperatures $\beta \in \left[0.25, 0.50, 0.75, 1.00, 1.50, 2.00\right]$ interpolated from $\beta=1.00$. Note that the sampler provides the positional information of the samples, so as before, we complement these samples by drawing their momentum components according to a Gaussian distribution with zero mean and co-variance $\frac{1}{\beta} \delta_{ij}$. Figure~\ref{fig:genai_histo} shows the histogram of the positional data obtained via TI model for $\beta \in \left[1.50, 2.00\right]$. 
One can see that the TI model produces samples whose distribution is very similar to the one of the rejection sampling scheme. The last column of the figure shows that results improve further after taking the importance weights of the samples into account, making the distributions near-indistinguishable.

\begin{figure}[ht]
    \centering
    \includegraphics[width=0.4\linewidth]{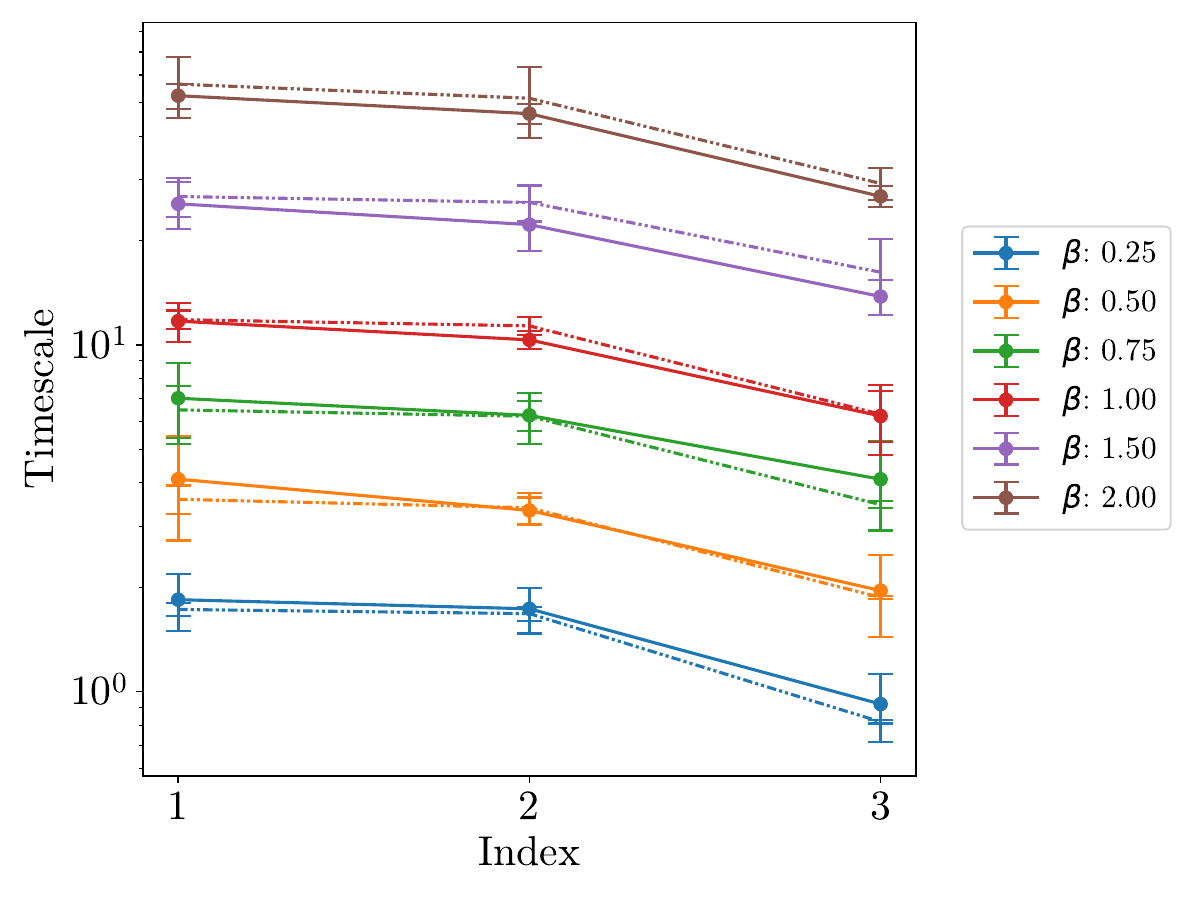}
    \caption{Slowest timescales $t_1$ to $t_3$ of the Lemon slice potential corresponding to different values of the inverse temperature $\beta$, computed using data generated via simulations of the underdamped Langevin SDE (solid lines) and via the TI generative model (dash-dot-dotted lines).}
    \label{fig:genai_com}
\end{figure}

Using the augmented TI dataset, we perform gEDMD as in Algorithm~\ref{alg:gEDMD_1} and compute the timescales of the system via estimation of the generator. Figure~\ref{fig:genai_com} shows the timescales corresponding to the temperature values listed above. Here we incorporated the sample weights computed via importance sampling only for $\beta \in \left[1.50, 2.00\right]$ as it does not affect other cases which were included in the training phase.

The figure shows that dataset generated via TI allows to recover the timescales computed via other sources of data up to statistical error. Thus, the TI data allows to predict the kinetic properties of a system that takes an order of magnitude longer to equilibrate relative to the training data. It is worth highlighting that the information about the case with $\beta=1.50$ has neither been involved in the training nor validation of the TI model, which emphasizes the extrapolation capabilities of TI. This is a significant result since the sole usage of the sampler to generate equilibrated samples takes considerably less time than integrating the SDE governing the underlying underdamped Langevin dynamics at unseen $\beta$. 

\subsection{CG dynamics learning}
\begin{figure}[ht]
    \centering
    \includegraphics[width=0.6\linewidth]{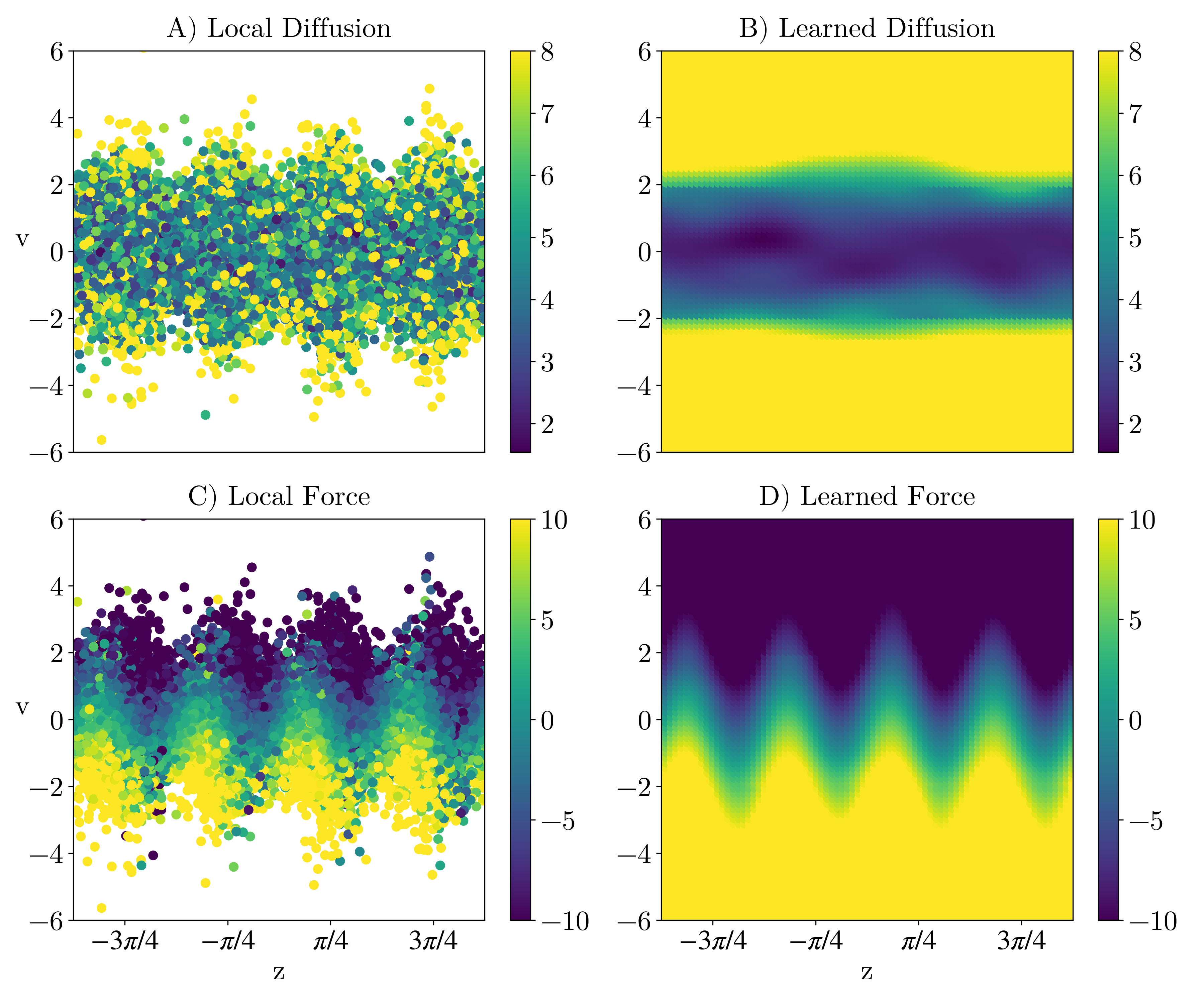}
    \caption{Learning of the effective diffusion ($\bar{\BA}_m$) and effective drift ($\bar{\bb}_m$) for $\beta=1$. \textbf{(A)} shows a scatter plot of the local diffusion $\BD_{\mathrm{loc}}$, while \textbf{(B)} shows the learnable component $\BD_m$ of the effective diffusion. Similarly, \textbf{(C)} shows a scatter of the local force $\bbf_{\mathrm{loc}}$, while \textbf{(D)} shows the learnable component $\bbf_m$ of the effective drift.}
    \label{fig:diff_learned_beta1}
\end{figure}

We now turn our attention to identifying the CG dynamics, by solving the two minimization problems~\eqref{eq:minimization_diff} in Section~\ref{subsec:lea_meth}. Figure~\ref{fig:diff_learned_beta1} shows the learned effective force and diffusion for $\beta = 1$. 
 
\begin{figure}[ht]
    \centering
    \includegraphics[width=0.6\linewidth]{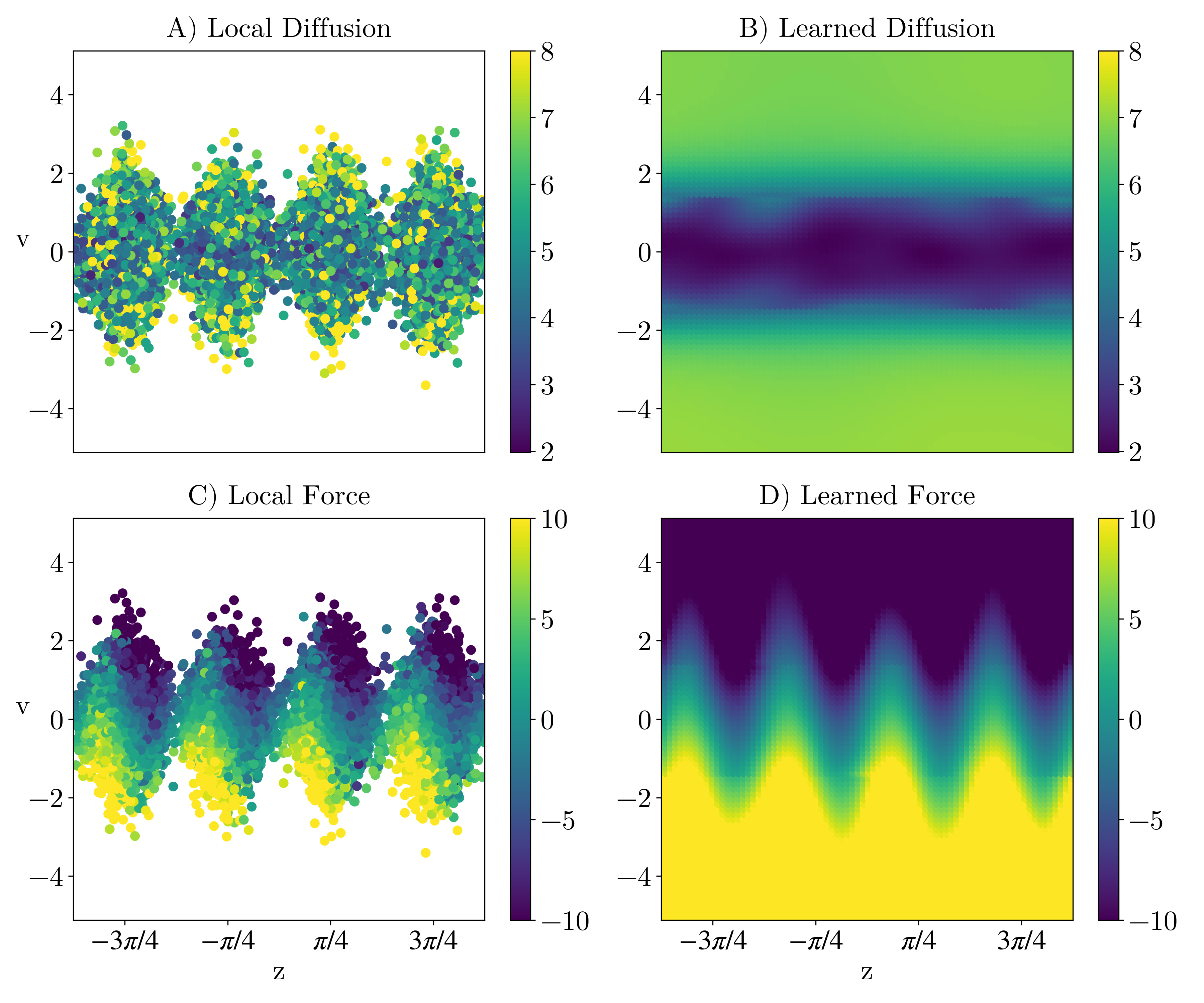}
    \caption{Learning of the effective diffusion ($\bar{\BA}_m$) and effective drift ($\bar{\bb}_m$) for $\beta=2$. \textbf{(A)} shows a scatter plot of the local diffusion $\BD_{\mathrm{loc}}$, while \textbf{(B)} shows the learnable component $\BD_m$ of the effective diffusion. Similarly, \textbf{(C)} shows a scatter of the local force $\bbf_{\mathrm{loc}}$, while \textbf{(D)} shows the learnable component $\bbf_m$ of the effective drift.}
    \label{fig:diff_learned_beta2}
\end{figure}

Results for learned diffusion show that it is almost constant as a function of $\bz$ and within the strongly populated region $\bv\in [-2, 2]$, while it keeps increasing in magnitude towards the tail of CG velocity distribution. A similar behavior can be observed for the learned drift, with the difference that it exhibits a sinusoidal behavior as a function of $\bz$. Note that the CG drift is two-dimensional, but we only need to learn the velocity component. We repeat the same process for the case of $\beta=2$, and it results in similar patterns as shown in Figure~\ref{fig:diff_learned_beta2}.

\subsection{Analysis of learned CG dynamics}
\begin{figure}[ht]
    \centering
    \begin{subfigure}[c]{0.4\textwidth}
    \includegraphics[width=\linewidth]{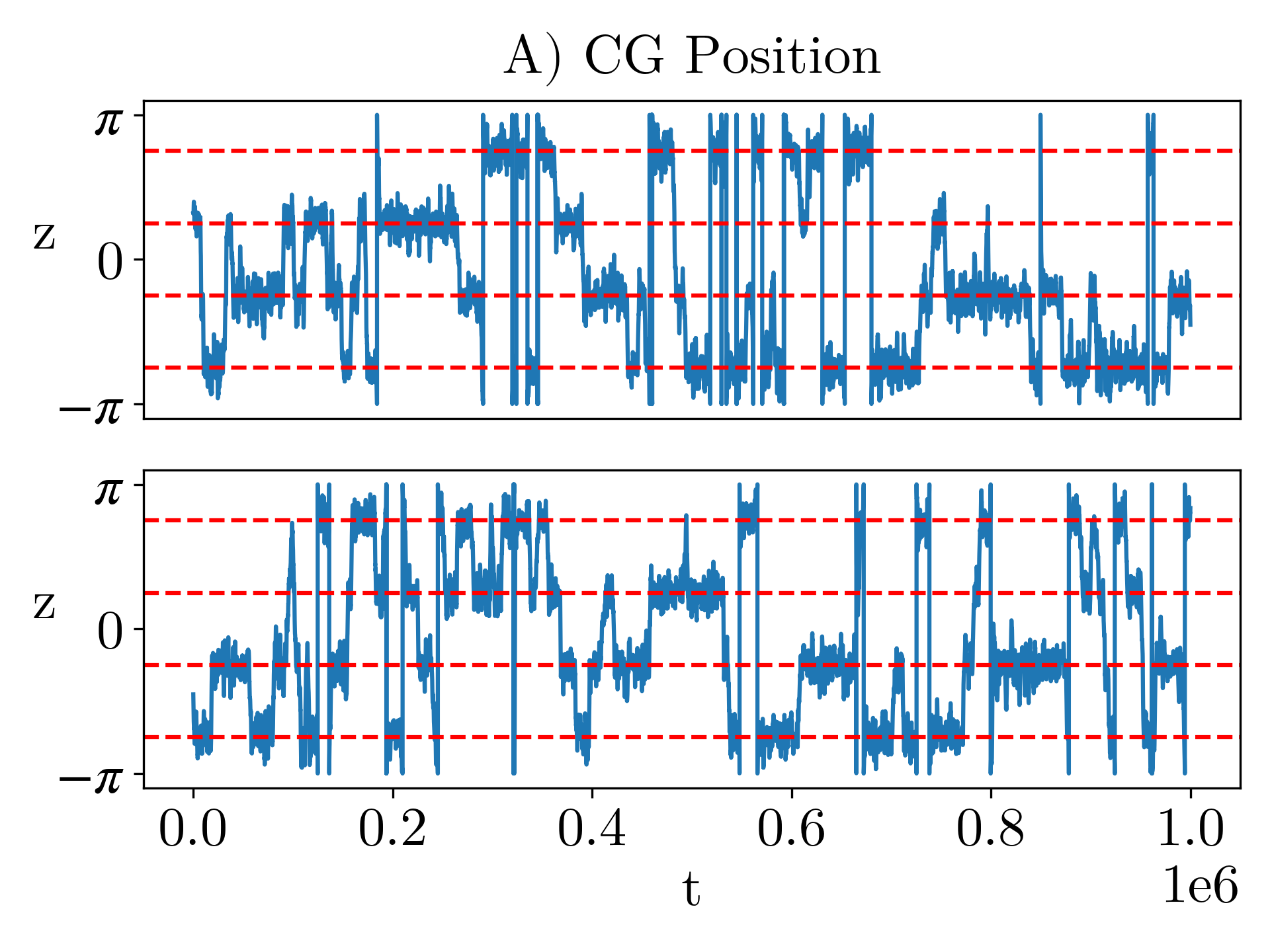}
    \end{subfigure}
    \begin{subfigure}[c]{0.4\textwidth}
    \includegraphics[width=\linewidth]{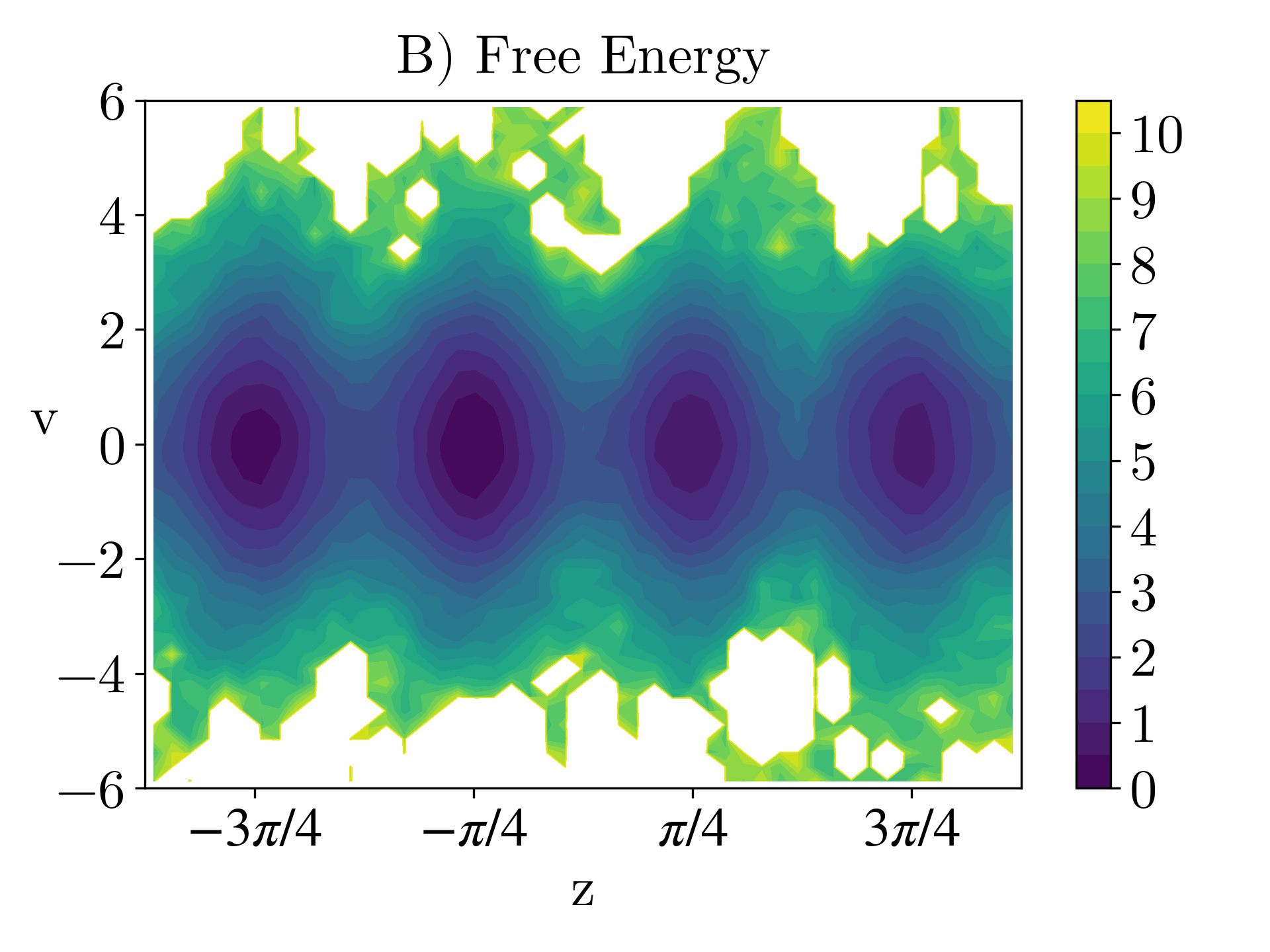}
    \end{subfigure}\\ 
    \begin{subfigure}[c]{0.4\textwidth}
    \includegraphics[width=\linewidth]{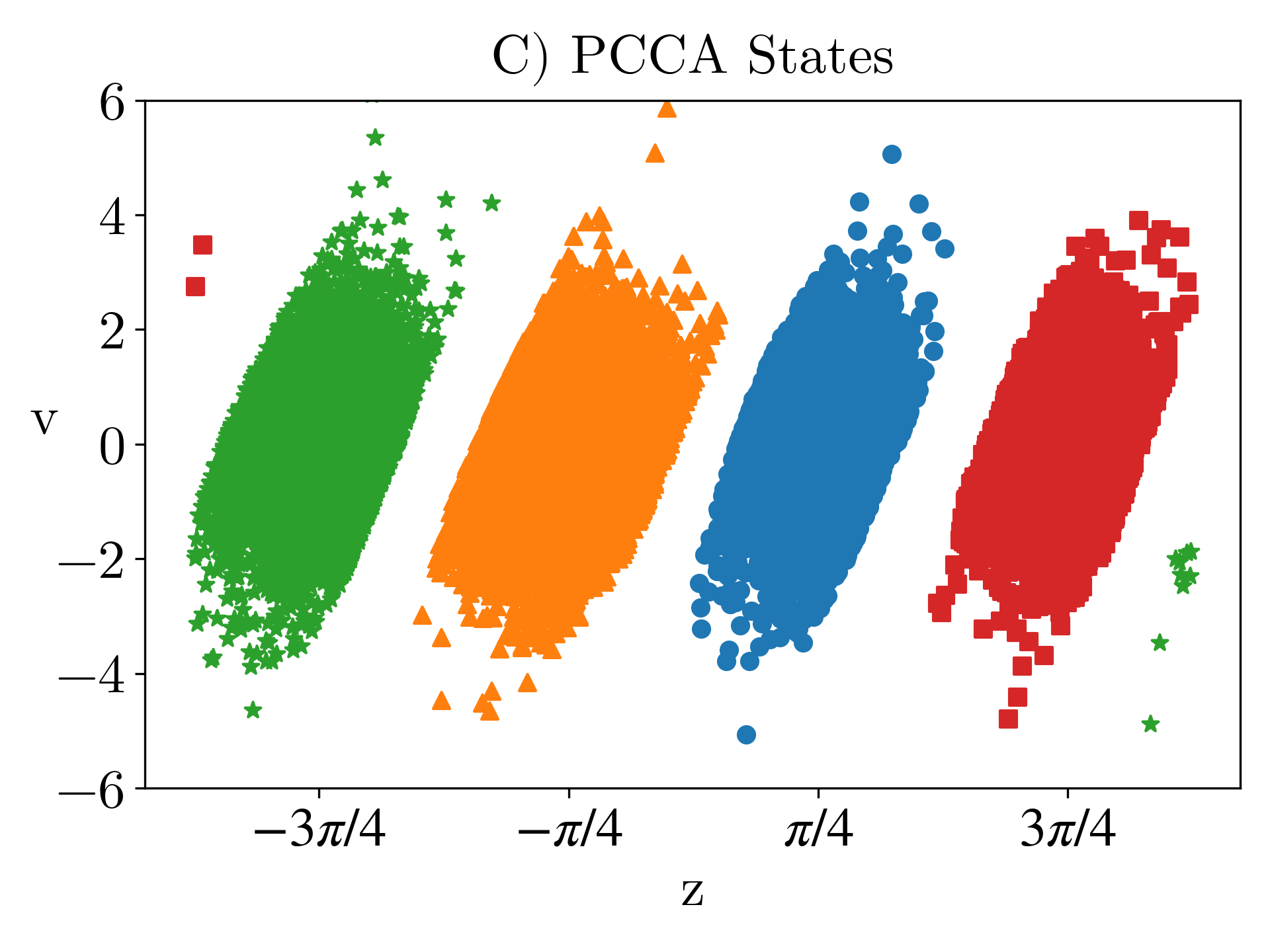}
    \end{subfigure}
    \begin{subfigure}[c]{0.4\textwidth}
    \includegraphics[width=\linewidth]{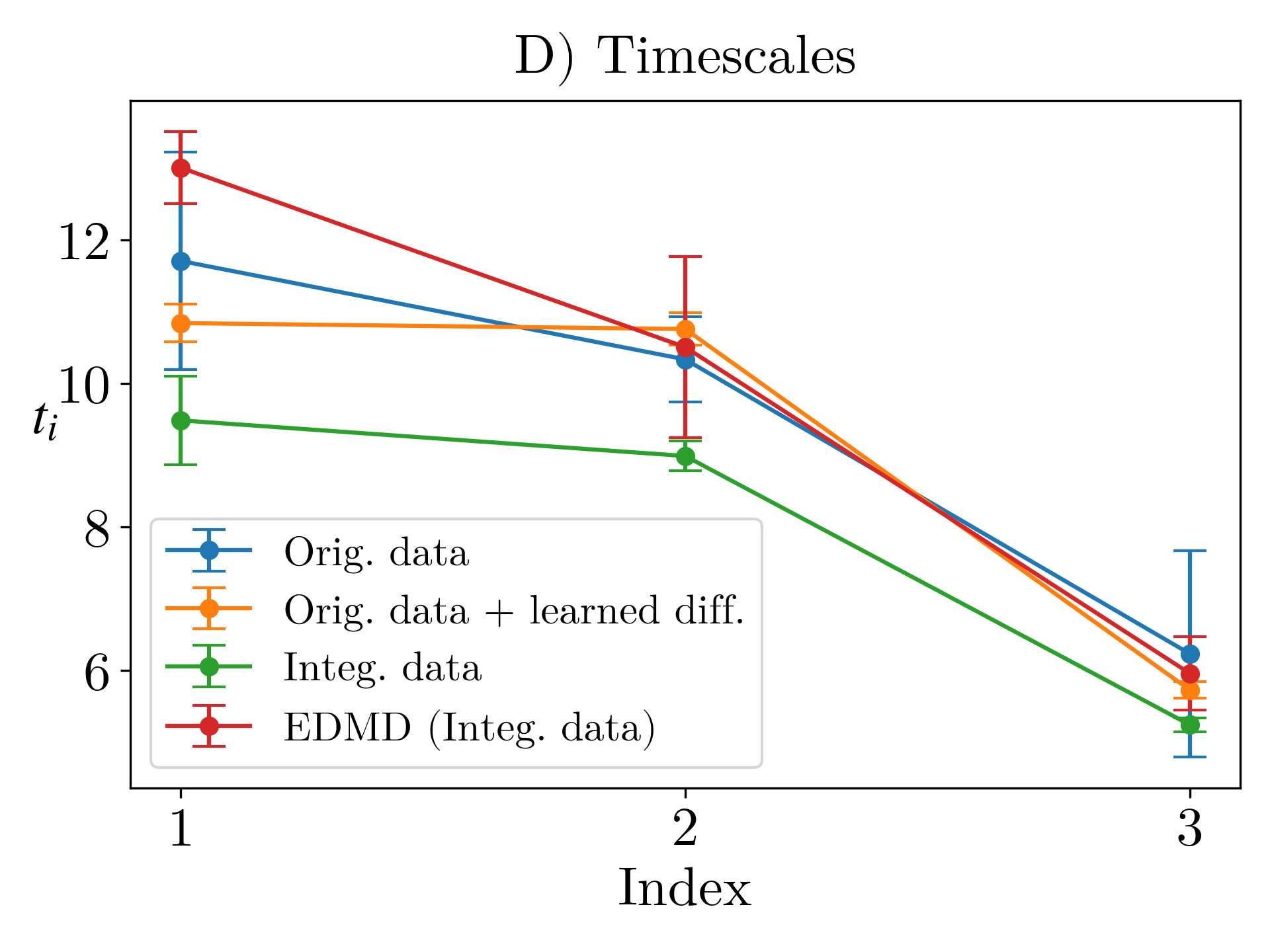}
    \end{subfigure}
    \caption{Integration of coarse grained SDE for $\beta=1$. Time evolution of the CG position for exemplary simulations in \textbf{(A)}, empirical free energy surface in \textbf{(B)}, PCCA+ analysis of the CG dynamics in \textbf{(C)}, and a comparison for the timescales of the coarse grained system in \textbf{(D)}. For a detailed description of panel \textbf{(D)}, see the main text.}
    \label{fig:sde_int_beta1}
\end{figure}

After learning the coefficients $\bar{\bb}_m$ and $\bar{\BA}_m$ of the CG Langevin dynamics, we integrate the learned dynamics to verify its statistical and dynamical properties agree with reference values. Figure~\ref{fig:sde_int_beta1}A shows the position space time series of two independent simulations of the CG dynamics at $\beta = 1$ for $10^6$ time steps. The trajectories visit the expected meta stable regions which are highlighted by horizontal dashed lines. We also compute an empirical free energy surface in the CG space based on these simulation data, as shown in Figure~\ref{fig:sde_int_beta1}B, which is in agreement with the potential of mean force obtained from the full dynamics.

In Figure~\ref{fig:sde_int_beta1}D, we compare multiple ways of assessing the slowest timescales of the CG Langevin process. First, we show the same timescales as in Figure~\ref{fig:its-comp} (blue lines), which were obtained by applying Algorithm~\ref{alg:gEDMD_1} to the original data in full space. These timescales represent a direct approximation of the CG generator $\cL^\Xi$ using the full space training data. Next, we can project the full space data into CG space, and then use the gEDMD method with these data and the learned parameters $\bar{\BA}_m$ and $\bar{\bb}_m$, to estimate a model for the generator of the learned CG dynamics. Note that this model for the learned CG generator, and its implied transition timescales, can be estimated without explicit integration of the CG Langevin dynamics. The corresponding implied timescales are shown by the orange line in Figure~\ref{fig:sde_int_beta1}D.

Using the CG simulation data and the gEDMD algorithm, we can compute yet another model for the learned generator in CG space. The corresponding timescales are depicted by the green line. Finally, we also apply the EDMD method at finite lag time to the CG simulation data, these timescales are represented by the red line. It is clear that all of these timescale estimates are in agreement within statistical error. These results show that we can evaluate transition timescales of the CG dynamics a priori by applying gEDMD to the full space training data. We can also assess the quality of the learned CG dynamics before running CG simulations. All of these results are consistent with the timescales we obtain after actually integrating the CG dynamics in time. 

We also apply robust Perron Cluster Cluster Analysis (PCCA+) algorithm~\cite{deuflhard_robust_2005} to the eigenvectors of the CG generator built upon the integrated dataset to assign to each sample point a degree of membership to each meta stable state. As depicted in Figure~\ref{fig:sde_int_beta1}C, we can confirm that meta stable states are also recovered. We conclude that all dynamical properties related to slow transitions are successfully retained by the learned dynamics.

\begin{figure}[ht]
    \centering
    \begin{subfigure}[c]{0.35\textwidth}
    \includegraphics[width=\linewidth]{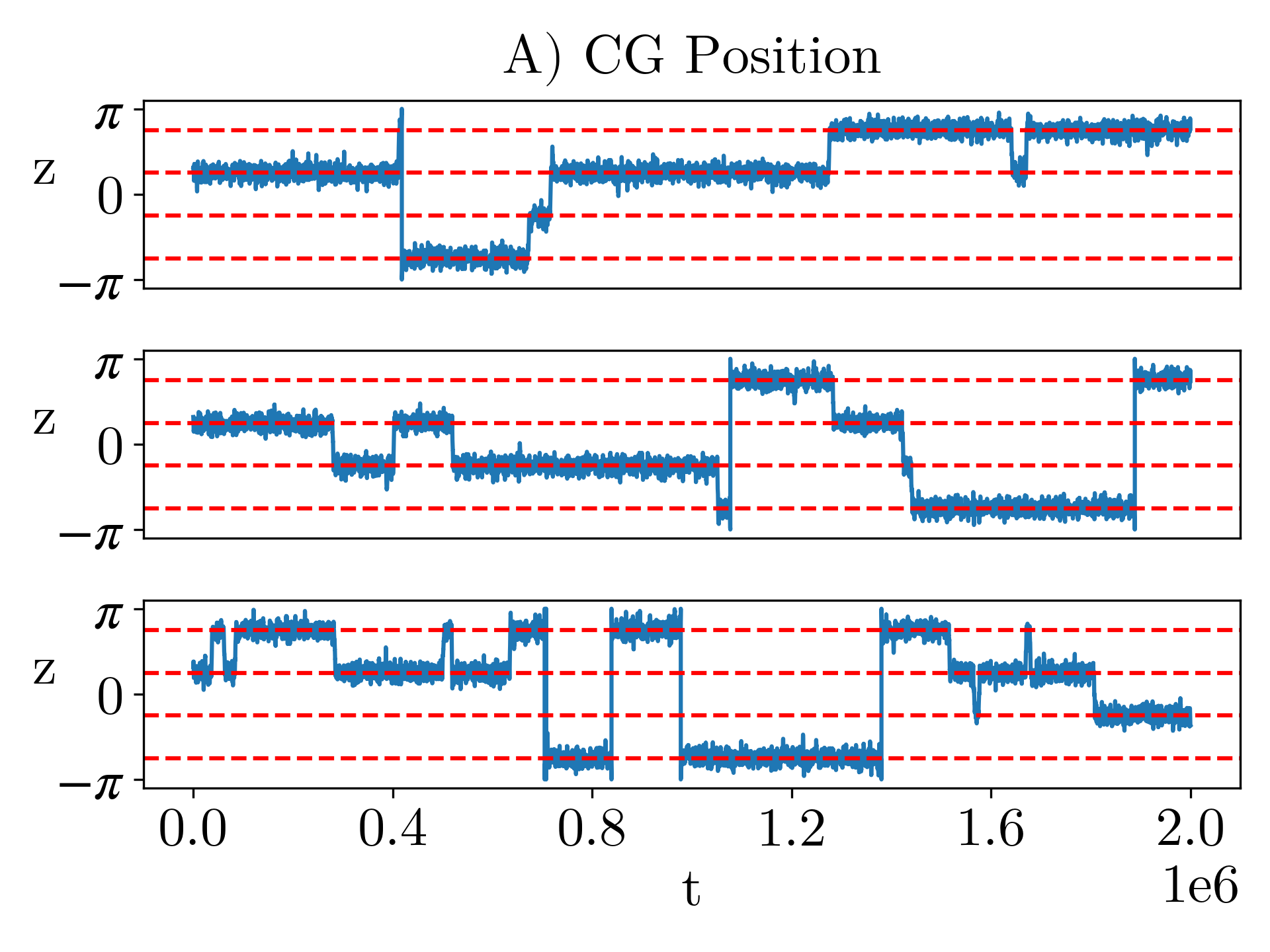}
     \end{subfigure}
    \begin{subfigure}[c]{0.35\textwidth}
    \includegraphics[width=\linewidth]{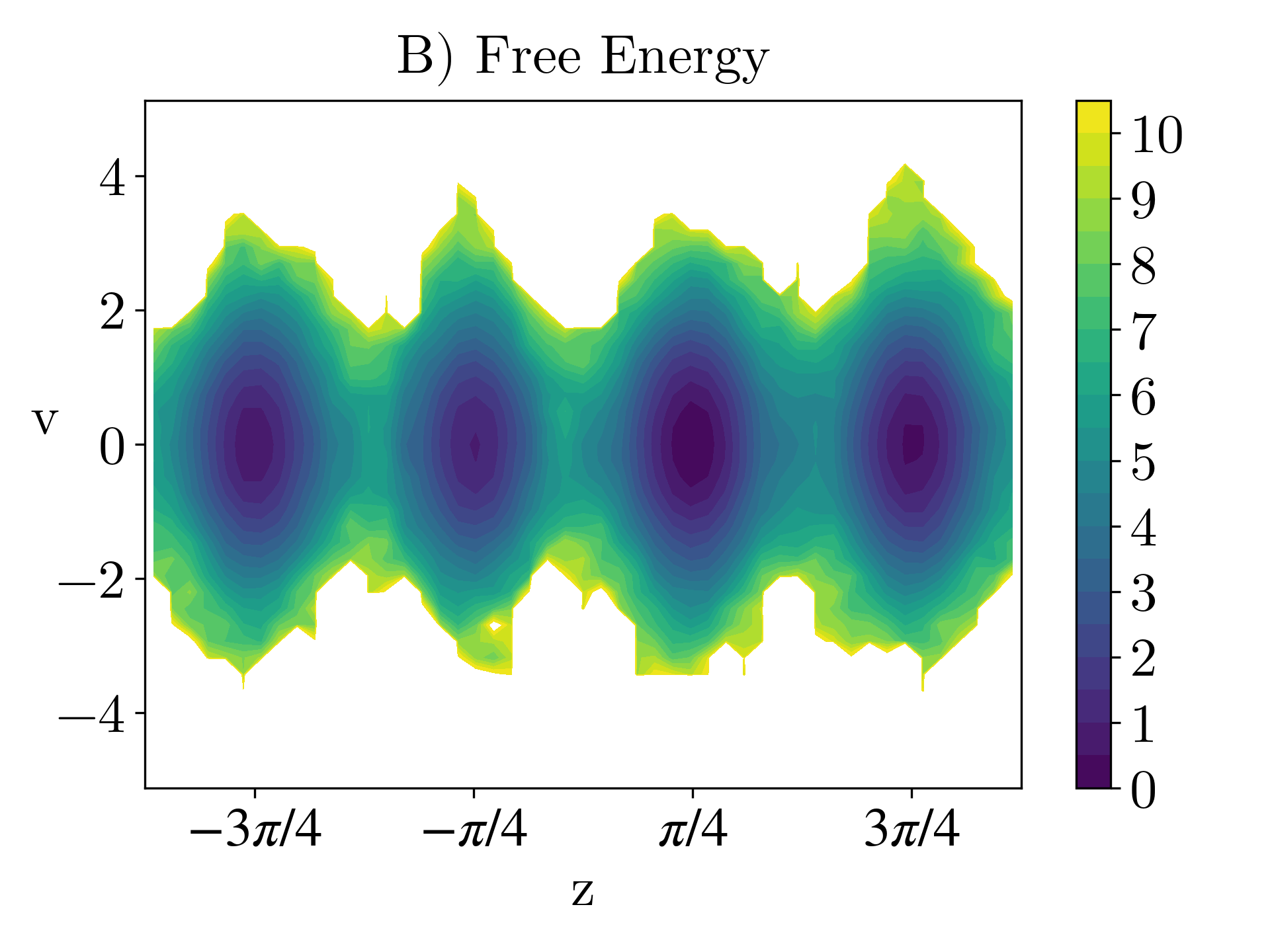}
    \end{subfigure}\\
    \begin{subfigure}[c]{0.35\textwidth}
    \includegraphics[width=\linewidth]{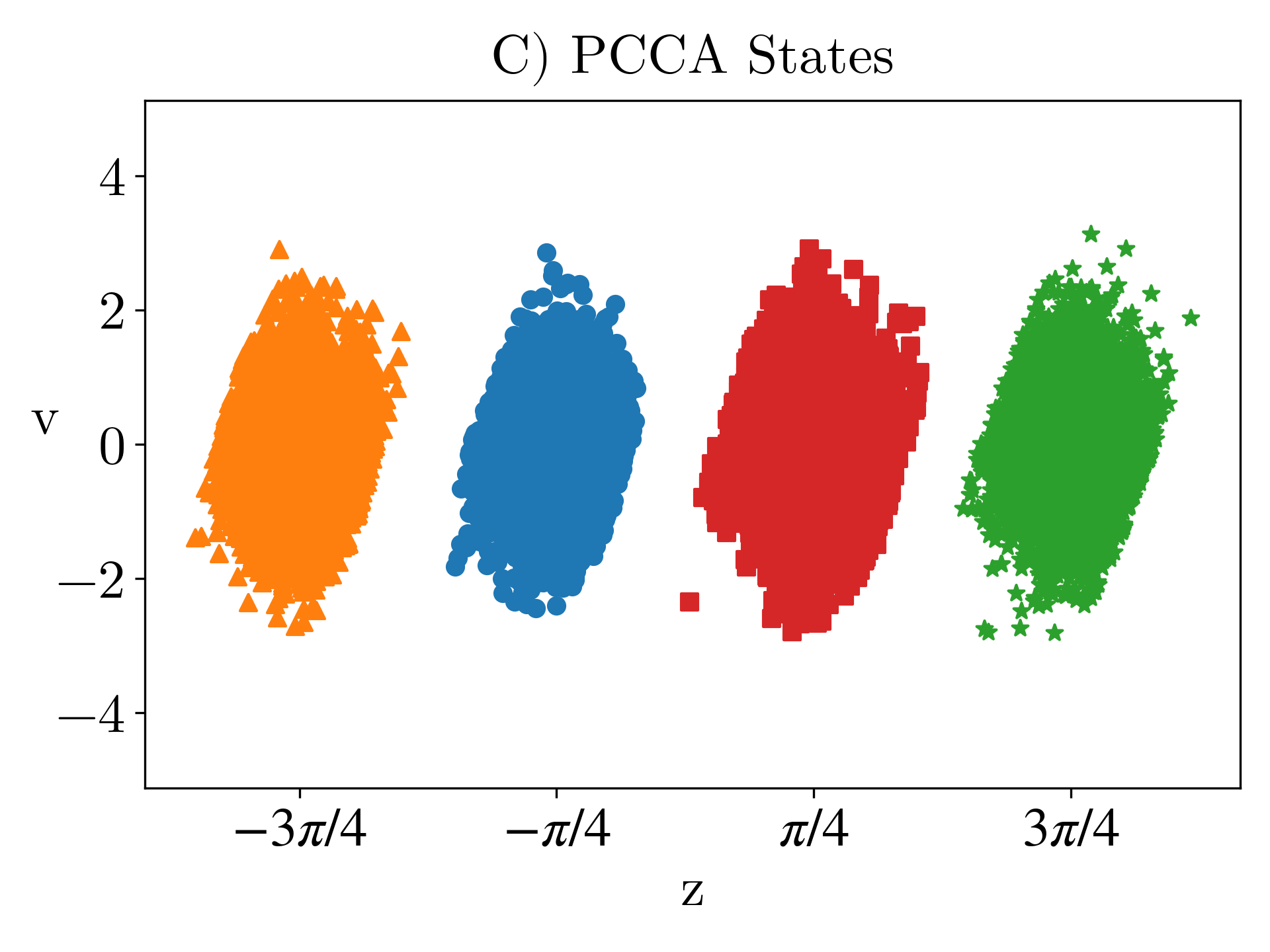}
    \end{subfigure}
    \begin{subfigure}[c]{0.35\textwidth}
    \includegraphics[width=\linewidth]{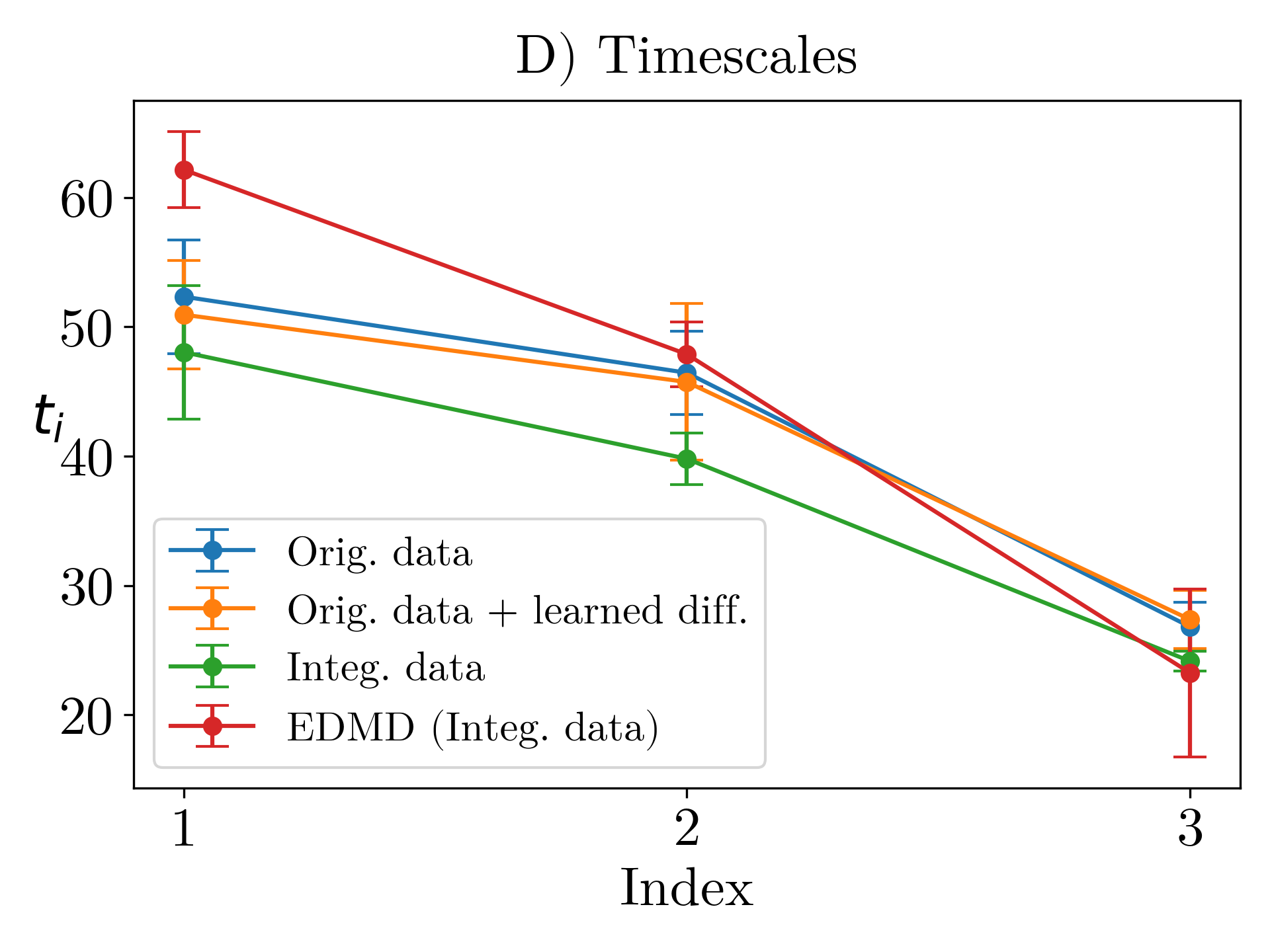}
    \end{subfigure}
    \caption{Integration of coarse grained SDE for $\beta = 2$. Time evolution of the CG position for exemplary simulations in \textbf{(A)}, empirical free energy surface in \textbf{(B)}, PCCA+ analysis of the CG dynamics in \textbf{(C)}, and a comparison for the timescales of the coarse grained system in \textbf{(D)}. For a detailed description of panel \textbf{(D)}, see the main text.}
    \label{fig:sde_int_beta2}
\end{figure}

Lastly, we repeat the same experiment for the case of $\beta = 2$. Figure~\ref{fig:sde_int_beta2} shows the results, which confirm all of our findings for the previous setting. We note that the CG dynamics are significantly more meta stable than for $\beta = 1$, as the oscillations around the minima are smaller and transitions occur rarely. In fact, extremely long CG simulations were needed to obtain statistically significant reference values. This highlights the value of our proposed approach: we can faithfully predict CG dynamics without long-time simulations, and their transition timescales without explicit integration of the CG dynamics. 

\section{Summary}
We investigated coarse graining of the underdamped Langevin dynamics based on a phase space CG map which uses both a coarse grained position and velocity component. We showed that application of the Zwanzig projection leads to a coarse grained dynamics with a similar structure as the full Langevin equation. We have also shown that the CG generator can be decomposed into symmetric and anti-symmetric parts resembling the Hamiltonian and Ornstein-Uhlenbeck parts of the full generator. Moreover, we showed that thermodynamic interpolation, a generative modeling technique that interpolates in temperature space, can be used to produce accurate samples of the position space distribution after training on high-temperature simulations. These samples can be combined with data-driven learning techniques, in particular the gEDMD method, to analyze kinetic properties of the CG dynamics, and to estimate its parameters, without explicit integration of the coarse grained model. By means of a two-dimensional model potential, we demonstrated that the CG dynamics reproduce thermodynamic and kinetic properties of the full-space dynamics.

\section*{Acknowledgments}
This work was partially supported by the Wallenberg AI, Autonomous Systems and Software Program (WASP) funded by the Knut and Alice Wallenberg Foundation (to S.O) and by a project grant from the Knut and Alice Wallenberg Foundation "From Atom to Organism: Bridging the scales in the design of ion channel drugs," (to S.M). The work of L.N. and C.H has been partially supported by the DFG Collaborative Research Center 1114 ``Scaling Cascades in Complex Systems'', project no. 235221301, project A05 ``Probing Scales in Equilibrated Systems by Optimal Nonequilibrium Forcing'' and by 
German Federal Government, the Federal Ministry of Research, Technology and Space and the State of Brandenburg within the framework of the joint project EIZ: Energy Innovation Center (project numbers 85056897 and 03SF0693A) with funds from the Structural Development Act (Strukturstärkungsgesetz) for coal-mining regions.

\section*{Data Availability}
The raw data as well as all code that are developed to generate the results are publicly available at~\cite{nateghiConsistentProjectionLangevin2026}.
\newpage
\appendix

\section{Proofs for Analytical Results}
\label{sec:proofs}
In this section, we provide technical proofs for Propositions~\ref{prop:CGdyn} and~\ref{prop:invmeas-gendecomp-CGdyn}. We start by verifying the structure of the CG parameters~\eqref{eq:force_diff_cg}, which follows quite directly after calculating the derivatives of the phase space CG map. The proof of Proposition 2 is along the lines of Proposition 4 in~\cite{zhang_reliable_2017}, which deals with CG for SDEs with non-degenerate noise. We spell out these results for the underdamped Langevin equation. In a few places, we use the summation convention that repeated indices are being summed over, as well as the shorthands $\partial_{i} = \frac{\partial}{\partial \bq_i},\ \partial_{ij} = \frac{\partial^2}{\partial \bq_i\partial \bq_j}$. 

\subsection{Proof of Proposition~\ref{prop:CGdyn}}
 Let us compute the derivatives of the CG map $\Xi$. We have for any $l \in \left\{ 1,\ldots,k\right\}$

\begin{align*}
    \nabla \bz_l &= \begin{pmatrix}
        \nabla \xi_l^\top \\ 0_d
    \end{pmatrix} \in \R^{2d}, \quad &&\nabla^2 \bz_l = \begin{pmatrix}
        \nabla^2 \xi_l & 0_{d\times d} \\ 0_{d\times d} & 0_{d\times d}
    \end{pmatrix} \in \R^{2d \times 2d}\\
     \nabla \bv_l &=  \begin{pmatrix}
        \nabla^2 \xi_l \BM^{-1} \bp \\ \nabla \xi_l^\top \BM^{-1}
    \end{pmatrix} \in \R^{2d}, \quad &&\nabla^2 \bv_l = \begin{pmatrix}
        \partial_{ijk} \xi_l m_i \bp_i & \nabla^2 \xi_l \BM^{-1} \\ \nabla^2 \xi_l \BM^{-1} & 0_{d\times d}
    \end{pmatrix} \in \R^{2d \times 2d}.
\end{align*}

The key observation is that the lower right blocks of the Hessians of all CG variables $\Xi_j$ vanish, which means that $\BA : \nabla^2\Xi_j \equiv 0$ for any $j \in \left\{1,\ldots,2k\right\}.$ Therefore, all second order terms vanish when computing the action of the generator $\cL \Xi$ on the phase space CG map. With this in mind, the computation of the effective drift $\bar{\bb}$ according to Eq.~\eqref{eq:cg-param} simplifies to:
\begin{align}
    \bar \bb (\bz,\bv) &= \cP(\cL \Xi) = \cP(\nabla \Xi \bb + 0) \\ 
    &= \cP \left(  \begin{pmatrix}
        \nabla \xi(\bq) & 0_{k\times d} \\ \bp^\top \BM^{-1} \partial^2 \xi(\bq) & \nabla \xi(\bq) \BM^{-1}
    \end{pmatrix} \cdot \begin{pmatrix}
        \BM^{-1}\bp \\ - \nabla V(\bq) - \gamma \bp 
    \end{pmatrix} \right) \notag \\ 
    &= \begin{pmatrix}
        \cP\left(\nabla \xi(\bq) \BM^{-1}\bp\right) \\ \cP\left( - \nabla \xi(\bq) \BM^{-1}\nabla V(\bq)  +  \bp^\top \BM^{-1} \partial^2 \xi(\bq) \BM^{-1}\bp - \gamma \nabla \xi(\bq) \BM^{-1}\bp \right) \end{pmatrix} \notag \\
        &= \begin{pmatrix}
            \bv \\ 
            \bbf(\bz, \bv) - \gamma \bv
        \end{pmatrix}.  \label{eq:eff-drift}
\end{align}

\noindent A similar calculation shows that the effective diffusion coefficient $\bar\BA$ is given by
\begin{align*}
   \bar \BA := \cP (\nabla \Xi \BA \nabla \Xi^\top) = \begin{pmatrix}
        0_{k\times k} & 0_{k\times k} \\ 0_{k\times k} & \gamma \cP(\nabla \xi(\bq) \BM^{-1}\nabla \xi(\bq)^\top)
    \end{pmatrix} = \begin{pmatrix}
        0_{k\times k} & 0_{k\times k} \\ 0_{k\times k} & \gamma \BD(\bz, \bv)
    \end{pmatrix},
\end{align*}
with $\BD(\bz, \bv)$ defined as $\BD(\bz, \bv) = \cP(\nabla \xi(\bq) \BM^{-1}\nabla \xi(\bq)^\top)(\bz, \bv)$.

\subsection{Proof of Proposition~\ref{prop:invmeas-gendecomp-CGdyn}}
The decomposition~\eqref{eq:gen_cg} of the coarse grained generator $\cL^\Xi$ is obtained in multiple steps. 

\paragraph{A different expression for the effective drift:}
First, we derive an alternative expression for the effective drift $\bar{\bb}$, which is given by the projection of the action of $\cL$ on the phase space CG map. We derive this relation in a weak form, that is, we  multiply $\bar{\bb}$ by an arbitrary function $h$ on the CG space, and integrate against the Boltzmann factor $e^{-\beta F}$. An application of the co-area formula~\eqref{eq:co_area_formula} together with the relation~\eqref{eq:prop_full_gen} then lead us to the following expression in terms of integrals on full phase space:
\begin{align*}
        \int_{\R^{2k}} & \bar{\bb}(\by) h(\by) e^{-\beta F(\by)} \diff \by\ 
        = \int_{\R^{2k}} \cP(\cL \Xi)(\by) h(\by) e^{-\beta F(\by)} \diff \by \\
        &= \int_{\R^{2d}} \cL \Xi(\bx) h(\Xi(\bx)) e^{-\beta H(\bx)} \diff \bx \\
        &= \innerprod{\jstat \nabla \Xi^\top}{ h \circ \Xi}_\mu - \frac{1}{\beta}\innerprod{\BA \nabla \Xi^\top}{ \nabla (h \circ \Xi)}_\mu.
\end{align*}
The gradient $\nabla(h \circ \Xi)$ is taken with respect to the full phase space variables $(\bq, \,\bp)$ and by the chain rule we have $\nabla_\bx (h\circ \Xi) = \nabla_\by h \, \nabla \Xi$. Inserting this relation, and applying the co-area formula in the reverse direction, we re-express the above in terms of CG space integrals and projections:
\begin{equation*}
    \begin{split}
        &\innerprod{\jstat \nabla \Xi^\top}{ h \circ \Xi}_\mu - \frac{1}{\beta}\innerprod{\BA \nabla \Xi^\top}{ \nabla (h \circ \Xi)}_\mu \\
        &= \int_{\R^{2k}} \cP(\jstat \nabla \Xi^\top)(\by) h(\by) e^{-\beta F(\by)} \diff \by - \frac{1}{\beta} \int_{\R^{2k}} \cP(\nabla \Xi \BA \nabla \Xi^\top)(\by) \nabla_\by h(\by) e^{-\beta F(\by)} \diff \by.
    \end{split}
\end{equation*}
It remains to apply a partial integration to the second term, which provides:
\begin{equation*}
\begin{split}
    \int_{\R^{2k}} & \bar{\bb}(\by) h(\by) e^{-\beta F(\by)} \diff \by \\
    &= \int_{\R^{2k}} \left[ \cP(\jstat \nabla \Xi^\top)(\by) + \frac{1}{\beta} e^{\beta F(\by)}\, \nabla_\by\cdot \left[ \cP(\nabla \Xi \BA \nabla \Xi^\top)(\by) e^{-\beta F(\by)}\right] \right] h(\by) e^{-\beta F(\by)} \diff \by.
\end{split}
\end{equation*}
As the function $h$ is arbitrary, and recognizing the definition~\eqref{eq:cg-param} of the effective diffusion field, we have shown that
\begin{equation}
\label{eq:decomp_cg_drift_prelim}
    \bar{\bb}(\by) = \cP(\jstat \nabla \Xi^\top)(\by) + \frac{1}{\beta} e^{\beta F(\by)}\, \nabla_\by\cdot \left[ \bar{\BA}(\by) e^{-\beta F(\by)}\right].
\end{equation}
Using that only the lower right block of $\bar{\BA}$ is non-zero, the second term simplifies to:
\begin{equation*}
    \frac{1}{\beta} e^{\beta F(\by)}\, \nabla_\by\cdot \left[ \bar{\BA}(\by) e^{-\beta F(\by)}\right] = \frac{\gamma}{\beta} e^{\beta F(\by)} \begin{pmatrix}
        0 \\
        \nabla_\bv \cdot \left[ \BD(\by) e^{-\beta F(\by)}\right]
    \end{pmatrix}.
\end{equation*}
This furnishes the final expression for the effective drift:
\begin{equation}
\label{eq:decomp_cg_drift}
    \bar{\bb}(\by) = \cP(\jstat \nabla \Xi^\top)(\by) + \frac{\gamma}{\beta} e^{\beta F(\by)} \begin{pmatrix}
        0 \\
        \nabla_\bv \cdot \left[ \BD(\by) e^{-\beta F(\by)}\right]
    \end{pmatrix}.
\end{equation}

\paragraph{A first decomposition of the effective generator}
Using Eq.~\eqref{eq:decomp_cg_drift}, we can start from the general definition of the CG generator $\cL^\Xi$, and re-write it as follows:
\begin{equation*}
    \begin{split}
        \cL^\Xi &= \bar{\bb} \cdot \nabla_\by + \frac{1}{\beta}\bar{\BA}:\nabla^2_\by \\
        &= \cP(\jstat \nabla \Xi^\top) \cdot \nabla_\by + \frac{\gamma}{\beta} e^{\beta F(\by)} \nabla_\bv \cdot \left[ \BD(\by) e^{-\beta F(\by)}\right] \cdot \nabla_\bv + \frac{\gamma}{\beta} \BD(\by) : \nabla^2_\bv.
    \end{split}
\end{equation*}
To re-write the last term, we have used again that only the lower right block of $\bar{\BA}(\by)$ is non-zero. The second and third term can be summarized as follows:
\begin{equation}
\label{eq:first_decomp_cg_gen}
    \cL^\Xi = \cP(\jstat \nabla \Xi^\top) \cdot \nabla_\by + \frac{\gamma}{\beta}  e^{\beta F(\by)} \nabla_\bv \cdot \left[ \BD(\by) e^{-\beta F(\by)} \nabla_\bv\right].
\end{equation}
The decomposition~\eqref{eq:first_decomp_cg_gen} is intriguing as the second term is in divergence form, which implies that it is automatically symmetric with respect to the inner product defined by $e^{-\beta F}$. The first term remains as a candidate for the anti-symmetric part, which is going to be verified next.

\paragraph{Continuity Equation}
Re-arranging Eq.~\eqref{eq:decomp_cg_drift_prelim}, we see that
\begin{equation*}
    \begin{split}
        \cP(\jstat \nabla \Xi^\top)(\by) &= \bar{\bb}(\by) - \frac{1}{\beta} e^{\beta F(\by)}\, \nabla_\by\cdot \left[ \bar{\BA}(\by) e^{-\beta F(\by)}\right],
    \end{split}
\end{equation*}
which is exactly the definition of the \emph{probability flow} associated to the Boltzmann factor $e^{-\beta F}$ under the CG dynamics~\cite{zhang_effective_2016}. If we can show that $\cP(\jstat \nabla \Xi^\top)$ satisfies the continuity equation
\begin{align*}
    - \nabla_\by \cdot \left[\cP(\jstat \nabla \Xi^\top)(\by) e^{-\beta F(\by)}\right] = 0,
\end{align*}
then it follows that $e^{-\beta F}$ is the invariant distribution for the CG dynamics. The continuity equation follows from a similar sequence of arguments that led us to the decomposition~\eqref{eq:decomp_cg_drift}. Again, we test against an arbitrary function $h$ and integrate over CG space:

\begin{equation*}
    \begin{split}
        - \int_{\R^{2k}} & \nabla_\by \cdot \left[\cP(\jstat \nabla \Xi^\top)(\by) e^{-\beta F(\by)}\right]\,h(\by)\,\diff \by = \int_{\R^{2k}}  \cP(\jstat \nabla \Xi^\top)(\by) e^{-\beta F(\by)} \nabla_\by \,h(\by)\,\diff \by \\
        &= \int_{\R^{2d}} \jstat(\bx) \nabla \Xi^\top(\bx) \nabla_\by h(\Xi(\bx)) e^{-\beta H(\bx)}\,\diff \bx,
    \end{split}
\end{equation*}
where we applied partial integration and the co-area formula. Using the chain rule and another partial integration, we arrive at the continuity equation in full space, which is zero as $\jstat$ is already the stationary probability for the full system:
\begin{equation*}
    \begin{split}
        \int_{\R^{2d}} & \jstat(\bx) \nabla \Xi^\top(\bx) \nabla_\by h(\Xi(\bx)) e^{-\beta H(\bx)}\,\diff \bx = \int_{\R^{2d}} \jstat(\bx) \cdot \nabla_\bx h(\Xi(\bx)) e^{-\beta H(\bx)}\,\diff \bx \\
        &= - \int_{\R^{2d}} \nabla_\bx\cdot \left[\jstat(\bx)e^{-\beta H(\bx)}\right] h(\Xi(\bx)) \,\diff \bx = 0.
    \end{split}
\end{equation*}
Indeed, $e^{-\beta F}$ is the invariant distribution of the CG dynamics, and we can introduce the notation
\begin{equation*}
    \bar{\BJ}^{\mathrm{eq}}(\by) := \cP(\jstat \nabla \Xi^\top)(\by)
\end{equation*}
for its stationary probability flow. The continuity equation also implies directly that the first term in Eq.~\eqref{eq:first_decomp_cg_gen} is anti-symmetric:
\begin{equation*}
    \begin{split}
        \int_{\R^{2k}} & \bar{\BJ}^{\mathrm{eq}}(\by) \cdot \nabla_\by \phi(\by) \,\psi(\by) e^{-\beta F(\by)}\,\diff \by \\
        &= - \int_{\R^{2k}} \nabla_\by\cdot \left[\bar{\BJ}^{\mathrm{eq}}(\by) e^{-\beta F(\by)}\right] \phi(\by)\, \psi(\by) \,\diff \by - \int_{\R^{2k}}  \bar{\BJ}^{\mathrm{eq}}(\by) \cdot \nabla_\by \psi(\by)\, \phi(\by) e^{-\beta F(\by)} \,\diff \by \\
        &= - \int_{\R^{2k}}  \bar{\BJ}^{\mathrm{eq}}(\by) \cdot \nabla_\by \psi(\by)\, \phi(\by) e^{-\beta F(\by)} \,\diff \by.
    \end{split}
\end{equation*}

\paragraph{Summary of the Results}
We have shown that Eq.~\eqref{eq:first_decomp_cg_gen} furnishes the decomposition of the CG generator into its symmetric and anti-symmetric parts, with invariant distribution $e^{-\beta F}$ and stationary probability flow $\bar{\BJ}^{\mathrm{eq}}$:
\begin{align*}
    \cL^{\Xi} &= \cL^{\Xi}_a + \cL^{\Xi}_s, \\
    \cL^{\Xi}_a \phi &= \bar{\BJ}^{\mathrm{eq}} \cdot \nabla_\by \phi, \\
    \cL^{\Xi}_s \phi &= \frac{\gamma}{\beta} e^{\beta F(\by)} \nabla_\bv \cdot \left[ \BD(\by) e^{-\beta F(\by)} \nabla_\bv \phi \right].
\end{align*}
Lastly, we note that $\bar{\BJ}^{\mathrm{eq}}$ indeed accounts for the Hamiltonian part of the CG drift:
\begin{align*}
        \bar{\BJ}^{\mathrm{eq}} &= \cP (\nabla \Xi \jstat) = \cP \left(  \begin{pmatrix}
            \nabla \xi(\bq) & 0_{k\times d} \\ \bp^\top \BM^{-1} \partial^2 \xi(\bq) & \nabla \xi(\bq)\BM^{-1}
        \end{pmatrix} \cdot \begin{pmatrix}
            \BM^{-1}\bp \\ - \nabla V(\bq)
        \end{pmatrix}\right) \\
       & = \begin{pmatrix}
            \bv \\
            \cP( \bp^\top \BM^{-1} \partial^2 \xi(\bq) \BM^{-1}\bp - \nabla \xi(\bq) \BM^{-1}\nabla V(\bq) )
        \end{pmatrix}
        = \begin{pmatrix}
            \bv \\
            \bbf 
        \end{pmatrix}.
    \end{align*}

\section{Simulation and Learning Details}
\subsection{Vamp-score Analysis}
\label{app:vamp}
\begin{figure}[ht]
    \centering
    \includegraphics[width=0.6\linewidth]{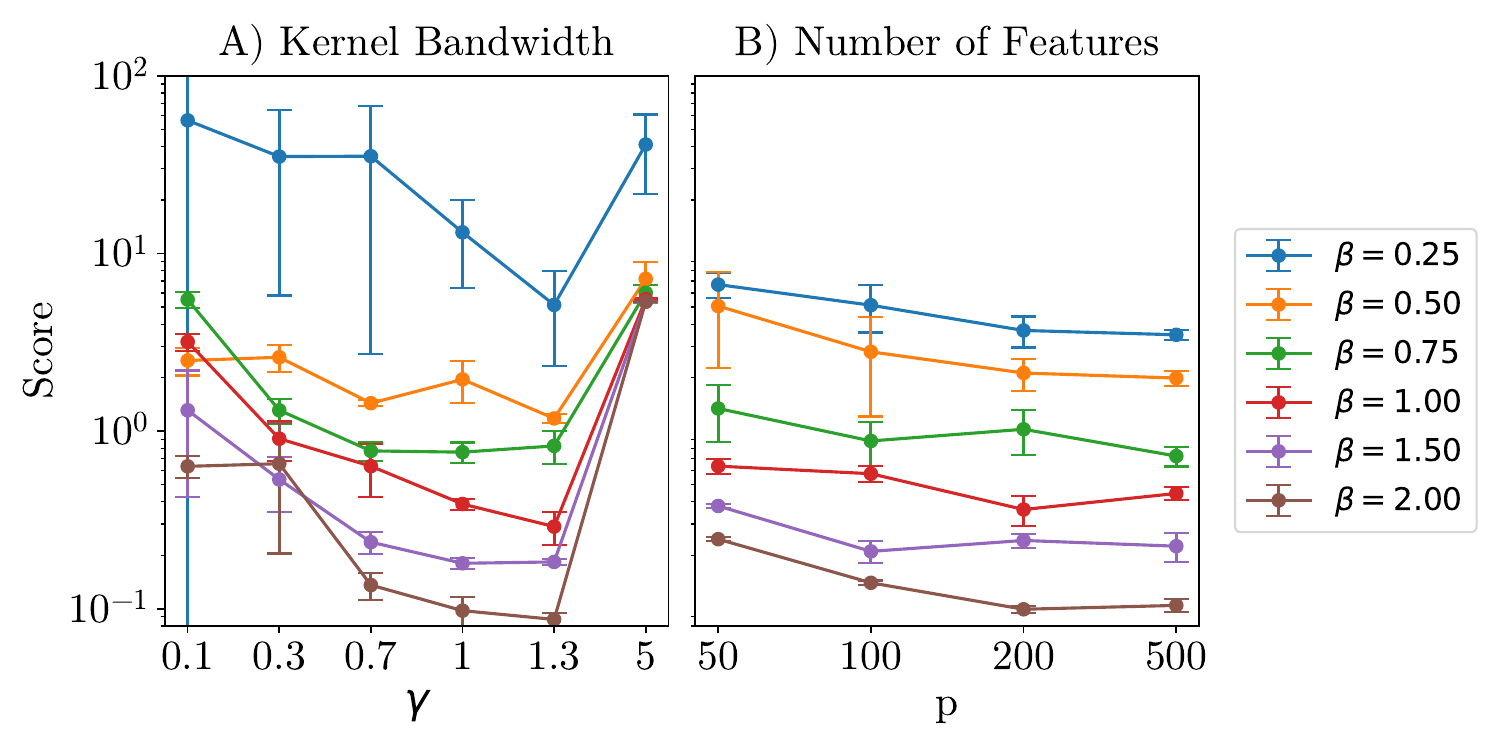}
    \caption{VAMP-score analysis. Score corresponding to various values of the kernel bandwidth and number of Fourier features in \textbf{(A)} and \textbf{(B)}, respectively.}
    \label{fig:vamp}
\end{figure}

We tune the parameters of the random Fourier feature basis for the approximation of the generator based on the VAMP variational principle proposed in Ref.~\cite{wu_variational_2020}. The two main parameters are the number of random Fourier features and the kernel bandwidth. Figure~\ref{fig:vamp} shows the result for these two parameters at given temperature. This leads to the following optimal parameters used to produce the timescale plots shown in the main text.

The VAMP-score analysis for the kernel bandwidth shows that there is a range of values that results in almost the same score, leading to flexibility in the choice of the parameter. For the number of features, we take the lowest value in the range where the score becomes constant. The summary of parameters relevant to the analysis is provided in Table~\ref{tab:rff_params}.  
\begin{table}[ht]
  \begin{center}
    \caption{Simulation setup}
    \begin{tabular}{|c|c|}
    \hline
        Kernel bandwidth ($\gamma$) & $1$ \\
        \hline
        Number of features ($p$) & $400$ \\
        \hline
        Tolerance in SVD truncation  & $10^{-4}$ \\
        \hline
      \end{tabular}
    \label{tab:rff_params}
  \end{center}
\end{table}
\subsection{Parameters for Langevin Simulations}
\label{app:sim_set}
We produced simulation data for the Lemon Slice system by integrating the Langevin SDE~\eqref{eq:Lan_underdamped} using Euler-Maruyama scheme, with the input setting summarized in Table~\ref{tab:setup}.
\begin{table}[ht]
  \begin{center}
    \caption{Simulation setup}
    \begin{tabular}{|c|c|}
    \hline
        Inverse temperature ($\beta$) & $\left[0.25,0.5,0.75,1.0,1.5,2.0\right]$ \\
        \hline
        Time constant ($1/\gamma$) & $0.2$ \\
        \hline
        Time step & $10^{-3}$ \\
        \hline
        Simulation time steps  & $10^6-10^7$ \\
        \hline
      \end{tabular}
    \label{tab:setup}
  \end{center}
\end{table}

\subsection{Neural network architecture for Parameters of the Effective Dynamics}
\label{app:nn_reg}
To learn the parameters $\bar{\bb}_m$ and $\bar{\BA}_m$ of the coarse grained dynamics, we adopt a shallow fully-connected feed-forward neural network with the setup summarized in Table~\ref{tab:nn_setup}.
\begin{table}[ht]
  \begin{center}
    \caption{Hyperparameters for the CG dynamics learning network}
    \begin{tabular}{|c|c|}
    \hline
        Batch size & $200$ \\
        \hline
        Number of hidden layer & $1$ \\
        \hline
        Hidden layer size & $8$ \\
        \hline
        Learning rate  & $\left(10^{-4}, 10^{-1}\right)$ \\
        \hline
        Optimizer  & Adam \\
        \hline
        Number of epochs & $200$ \\
        \hline
      \end{tabular}
    \label{tab:nn_setup}
  \end{center}
\end{table}

\subsection{TI}
\label{app:ti}
For the TI model we employ the two-sided linear interpolant with a sine-squared noise schedule \cite{albergo_stochastic_2023}, implying that the interpolant is defined as
\begin{equation}
    \bx(\tau) = (1 - \tau)\bx_0 + \tau\bx_1 + \sin^2(\pi \tau) \bz,
\end{equation}
for $\tau \in [0,\,1]$, $\bx_0\sim\rho_0(\bx_0)$, $\bx_1\sim\rho_1(\bx_1)$ and $\bz\sim\mathcal{N}(0,\,\mathrm{Id})$.

To encourage extrapolation of the TI model to temperatures unseen during training, we embed the inverse temperatures $\beta$ using a positional embedding $\lambda_\mathrm{pos}(\beta;l_\beta)$ \cite{moqvist_thermodynamic_2025}. Similarly, we also use positional embeddings $\lambda_\mathrm{pos}(\tau;l_\tau)$ and $\lambda_\mathrm{pos}(\bx(\tau);l_\bx)$ to embed the positions $\bx$ and times $\tau$. We define the embeddings for some model feature $\bx$ as $\lambda_\mathrm{pos}(\bx(\tau);l_\bx) = \{\lambda_\mathrm{pos}^n(\bx;l_\bx)\}_{n=0}^N$ where
\begin{equation}
    \lambda_\mathrm{pos}^n(\bx;l) = \begin{cases}
        \cos\left(\left(1 + \frac{n}{2}\right)+\frac{\pi\beta}{l}\right)\quad n\text{ even}\\
        \sin\left(\left(1 + \frac{n-1}{2}\right)+\frac{\pi\beta}{l}\right)\quad n\text{ odd},
    \end{cases}
\end{equation}
for some maximal embedding dimension $N$. Then, the embedded $\tau$, $\beta_0$, $\beta_1$ and $\bx(t)$ are projected through individual linear layers, concatenated and fed into a shared MLP. The network architecture and the corresponding hyperparameters are summarized in Table~\ref{tab:ti_setup}.
\begin{table}[ht]
  \begin{center}
    \caption{Hyperparameters and TI model settings.}
    \begin{tabular}{|c|c|}
    \hline
        Batch size & $1024$ \\
        \hline
        Number of layers in MLP & $5$ \\
        \hline
        Positional embedding size, N & $256$ \\
        \hline
        Hidden layer size & $256$ \\
        \hline
        Learning rate  & $10^{-4}$ \\
        \hline
        Optimizer  & Adam \\
        \hline
        Number of epochs & $75$ \\
        \hline
        Positional encoder scale, $l_\beta$ & $350$ \\
        \hline
        Positional encoder scale, $l_t$ & $10$ \\
        \hline
        Positional encoder scale, $l_\bx$ & $50$ \\
        \hline
        Numerical solver & Euler \\
        \hline
      \end{tabular}
    \label{tab:ti_setup}
  \end{center}
\end{table}

\pagebreak

\bibliographystyle{unsrt}
\bibliography{references,other}

\end{document}

%% file: commands.tex
\usepackage{amsmath,amsfonts,amssymb,amsthm}
\usepackage{todonotes}
\usepackage{xcolor}

\newcommand{\R}{\mathbb{R}}                                     

\newcommand{\innerprod}[2]{\left\langle #1,\, #2 \right\rangle} 

\newcommand{\diff}{\mathrm{d}}                                   

\DeclareMathOperator*{\argmin}{arg\,min}

\newcommand{\bE}{\mathbb{E}}

\newcommand{\cL}{\mathcal{L}}
\newcommand{\cP}{\mathcal{P}}

\newcommand{\BA}{\mathbf{A}}
\newcommand{\BB}{\mathbf{B}}

\newcommand{\BD}{\mathbf{D}}

\newcommand{\BG}{\mathbf{G}}

\newcommand{\BJ}{\mathbf{J}}

\newcommand{\BL}{\mathbf{L}}
\newcommand{\BM}{\mathbf{M}}

\newcommand{\BR}{\mathbf{R}}

\newcommand{\BS}{\mathbf{S}}

\newcommand{\BU}{\mathbf{U}}
\newcommand{\BV}{\mathbf{V}}
\newcommand{\BW}{\mathbf{W}}
\newcommand{\BX}{\mathbf{X}}
\newcommand{\BY}{\mathbf{Y}}

\newcommand{\bb}{\mathbf{b}}

\newcommand{\bbf}{\mathbf{f}}

\newcommand{\bq}{\mathbf{q}}
\newcommand{\bp}{\mathbf{p}}

\newcommand{\bv}{\mathbf{v}}

\newcommand{\bx}{\mathbf{x}}
\newcommand{\by}{\mathbf{y}}
\newcommand{\bz}{\mathbf{z}}

\newcommand{\jstat}{\mathbf{J}^{\mathrm{eq}}}

\newcommand{\jac}{\mathrm{Jac}}

\newtheorem{theorem}{Theorem}

\newtheorem{proposition}[theorem]{Proposition}